\documentclass[
 reprint,
 superscriptaddress,
 amsmath,amssymb,
 aps,
 prapplied,
]{revtex4-2}

\usepackage{graphicx}
\usepackage{dcolumn}      
\usepackage{bm}
\usepackage{hyperref}
\usepackage{caption}
\usepackage{subcaption}
\usepackage{ragged2e}
\usepackage{physics}
\usepackage{amsmath}
\graphicspath{{Images_submission/}}
\usepackage{float}
\usepackage[utf8]{inputenc}
\usepackage{textgreek}
\usepackage{csquotes}
\usepackage[normalem]{ulem}

\begin{document}

\title{Optimized Dynamical Decoupling for Spin Ensembles with Large Inhomogeneities}

\author{Rajan Paul}
\email{rajan.paul@uni-ulm.de}
\affiliation{Institute of Quantum Optics, Ulm University, Albert-Einstein-Allee 11, 89081 Ulm, Germany}
\author{Genko Genov}
\affiliation{Institute of Quantum Optics, Ulm University, Albert-Einstein-Allee 11, 89081 Ulm, Germany}
\author{Oliver Schmid}
\affiliation{Institute of Quantum Optics, Ulm University, Albert-Einstein-Allee 11, 89081 Ulm, Germany}

\author{Shinobu Onoda}
\affiliation{Quantum Materials and Applications Research Center, National Institutes for Quantum Science and Technology, Takasaki, 370-1292, Japan}
\author{Yutaka Kobayashi}
\affiliation{Advanced Materials Laboratory, Sumitomo Electric Industries Ltd., Itami, 664-0016, Japan}

\author{Junichi Isoya}
\affiliation{R\&D Center for Innovative Material Characterization, University of Tsukuba, Tsukuba, Ibaraki, 305-8550, Japan}
\author{Fedor Jelezko}
\affiliation{Institute of Quantum Optics, Ulm University, Albert-Einstein-Allee 11, 89081 Ulm, Germany}
\date{\today}

\begin{abstract}
Spin ensembles show increasing promise for highly efficient quantum sensing and quantum technologies. However, protecting their coherent properties remains a significant challenge. Dynamical decoupling (DD) typically employs a sequence of fast pulses that aim to reduce the effect of unwanted system-environment interactions. However, the performance of the DD sequences typically suffers in spin ensembles, e.g., due to amplitude and frequency inhomogeneities and control errors.  
In this work, we compare the performance of an extensive number of DD protocols for coherence protection and quantum sensing in an ensemble of negatively charged NV centers under ambient conditions.  
We identify the optimal approach that uses rapid adiabatic passage pulses, which are robust to amplitude and frequency inhomogeneity, in combination with the variation of their phases.  
We experimentally demonstrate an improved signal contrast and an order-of-magnitude increase in memory time with the optimal sequence compared to the widely used XY8 with standard rectangular pulses. We also apply our approach for improved sensing of a weak oscillating field. While our particular demonstration is with NV centers in diamond, the optimized protocol is generally applicable to other spin ensembles and to systems with high amplitude and/or frequency inhomogeneities.
\end{abstract}

\maketitle
\section{Introduction}
\label{sec:intro}
Quantum technologies \cite{dowling2003quantum, acin2018quantum} are becoming increasingly important nowadays for multiple applications in communication \cite{kimble2008quantum}, computation \cite{nielsen2001quantum}, simulation \cite{feynman2018simulating}, sensing \cite{degen2017}, and metrology \cite{Georgescu_2012}. Spin ensembles, including nitrogen-vacancy (NV) centers in diamond \cite{JelezkoPRL2004,balasubramanian2008nanoscale,maze_nanoscale_2008}, silicon carbide (SiC) \cite{castelletto_quantum_2023}, and hexagonal boron nitride (hBN) \cite{healey_quantum_2023} are promising experimental platforms for a wide range of such applications, and quantum sensing in particular \cite{degen2017,barry_sensitivity_2020}. The NV centers are especially useful as their electron spin can be initialized, manipulated, and read out under ambient conditions using electron spin resonance techniques \cite{balasubramanian2008nanoscale,DOHERTY20131,barry_sensitivity_2020}. The interaction of the NV spin states with external electromagnetic fields makes them particularly amenable to applications in spin-based quantum sensing, e.g., as a nanoscale sensor for temperature \cite{kucsko2013nanometre}, pressure \cite{doherty2014electronic}, strain \cite{trusheim2016wide}, electric \cite{dolde2011electric}, and magnetic fields \cite{balasubramanian2008nanoscale}.

\begin{figure*}[t]
\centering

\includegraphics[width=\textwidth]
{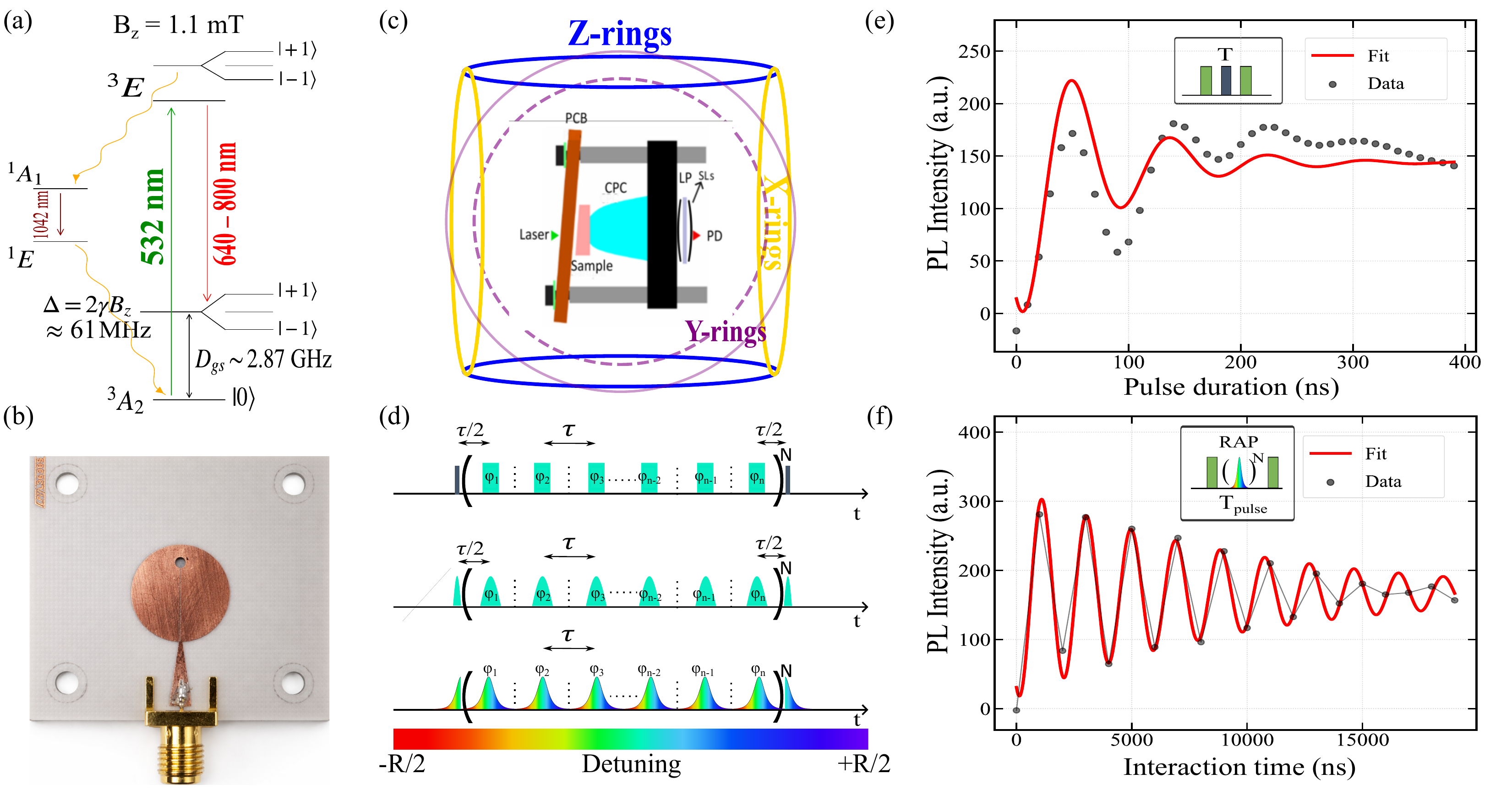}
\caption{(a) Energy level diagram of the NV$^{-}$ center at bias magnetic field of amplitude $1.1$ mT. The NV is prepared in state $|0\rangle$ by an off-resonant green laser ($532$ nm), coherently manipulated using microwave fields coupling states $|0\rangle$ and $|\text{-}1\rangle$, and optically read out via spin-dependent fluorescence \cite{JelezkoPRL2004,barry_sensitivity_2020}. (b) Home-built microwave resonator, labeled PCB, used in the experiment, modified from Sasaki et al. \cite{sasaki2016broadband}. (c) Cross-sectional schematic of the experimental assembly. The Helmholtz coil encloses the optical detection system. Labels: CPC, compound parabolic concentrator; SL, spherical lens; LP, long-pass filter; PD, photodiode. The pink element represents the diamond sample. (d) Schematic representation of dynamical decoupling (DD) with standard, shaped and rapid adiabatic pulses (RAPs). (Top) The standard measurement scheme with rectangular pulses. The experiment begins and ends with $\pi/2$ pulses (dark grey), while DD is performed with a series of $n$ refocusing $\pi$ pulses with phases $\phi_1,\dots\phi_n$. The time between the pulses' centers is $\tau$, with $\tau/2$ giving the time between the end (beginning) of the first (last) $\pi/2$ pulse and the  center of the first (last) $\pi$ pulse. (Middle) The same sequence with sin or parabola-shaped pulses.  (Bottom) RAP-based protocol, where the refocusing pulses are RAPs. The two $\pi/2$ pulses at the two ends are replaced by the half-RAPs, as indicated by the leading (lagging) edge pulses \cite{genov2020efficient}. The color gradient within each pulse corresponds to the change in the frequency detuning $\delta(t)$ from $-R/2$ (red) to $+R/2$ (blue) with a chirp range of $R$. (e) Rabi oscillations with standard pulses were observed to decay rapidly, with an initial contrast of $17\%$. Fitting the decay with a damped cosine (red curve),
$a\exp(-t/T_\Omega)\cos(2\pi\Omega t+\phi)+c$, yields a Rabi frequency of $\Omega\approx11.45$~MHz and a
decay time $T_\Omega$. The amplitude inhomogeneity is characterized by its standard deviation 
$\sigma_\epsilon=\sqrt{2}/(\Omega T_\Omega)$~\cite{Kitamura2026praaplied}, assuming the Gaussian-decay form
$\exp\{-(\Omega t\sigma_\epsilon)^2/2\}$. (inset) Pulse sequence for Rabi measurements: initialization with green laser pulse is followed by a resonant standard microwave pulse (dark grey), applied for duration $T$, and a subsequent laser pulse for readout. (f) RAP-based Rabi oscillations, obtained by replacing the continuous driving field for the standard Rabi measurement with RAPs of duration $T_{\text{pulse}}$, repeated multiple $1,\dots N$ times (see inset for a schematic of the sequence). The RAP-Rabi contrast improves to $24.5\%$, a relative increase of $44\%$ in comparison to the standard Rabi oscillations experiment, and prolongs the number of oscillations until the signal decays.  }
\label{fig:1}
\end{figure*}

While single NV centers are very suitable for high-spatial-resolution field imaging \cite{balasubramanian2008nanoscale}, they often exhibit limited magnetic sensitivity, which can be several orders of magnitude worse than the theoretical limits \cite{barry_sensitivity_2020}. This is typically due to the low readout efficiency of single NV centers, e.g., due to photon-shot noise \cite{chuprina2026impactphotoelectricreadoutnoise}. In contrast, the use of NV ensembles can significantly improve sensitivity \cite{barry2024sensitive}, as the signal-to-noise ratio usually scales as $\sqrt{N_{\text{NV}}}$ where $N_{\text{NV}}$ is the number of non-interacting identical NV centers present in the ensemble volume 
\cite{barry_sensitivity_2020,chuprina2026impactphotoelectricreadoutnoise}. This has allowed us to reach sensitivities of the order of several pT/$\sqrt{\mathrm{Hz}}$ \cite{PhysRevX.5.041001} and sub-pT/$\sqrt{\mathrm{Hz}}$ \cite{barry2024sensitive} with ensembles of NV centers.  

However, a significant remaining challenge is the limited coherence time $T_2$ in such experiments, e.g., due to amplitude, frequency inhomogeneity, and pulse errors \cite{barry_sensitivity_2020}. As the sensitivity typically scales as $\sim T_2^{-1/2}$ \cite{pham2011magnetic, pham2012enhanced,degen2017}, enhancing the coherence time remains an important challenge. To address this problem, one can typically employ dynamical decoupling (DD). DD consists of a sequence of fast, ideally instantaneous, $\pi$ pulses, which aim to nullify the effect of the interactions of the NV centers with their surrounding environment \cite{viola_dynamical_1999,viola_universal_1999}. However, due to many pulses and the inhomogeneities in case of spin ensemble experiments, the fidelity of the DD sequence can drop fast with the number of pulses. Although ensemble $T_2$ has been significantly improved with these methods at cryogenic temperatures \cite{de2010universal, bar2013solid}, reaching long coherence times at room temperature remains challenging in the presence of inhomogeneities. For example, Ramsey and simple Hahn echo experiments have been used to achieve optimal sensitivities in recent ensemble experiments \cite{barry2024sensitive} although their coherence times are typically much worse than the ones achievable by XY-based DD protocols \cite{Maudsley1986xy4,Gullion1990xy,farfurnik2015optimizing} for single NV centers \cite{shim2012robust, naydenov2011dynamical,barry_sensitivity_2020}. 
Specifically, the coherence time usually scales as $T_2\sim N^{2/3}$ with $N$ the number of ideal $\pi$ pulses in a sequence \cite{bar2013solid,romach2015spectroscopy,senkalla2024germanium,grimm2025coherent}.  
However, in the case of large inhomogeneities, pulse imperfections could accumulate during the sequence, leading to lower coherence times for higher $N$ \cite{wang2012effect}.

In this work, we investigate an extensive number of DD protocols \cite{Maudsley1986xy4,Gullion1990xy,souza2011robust,Souza2012,genov2017arbitrarily,wang2019randomization,Ezzell23} - including composite pulses \cite{levitt_composite_1986,wimperis1994broadband,Torosov2011smooth,TorosovPRL2011,genov2014correction}, and Hamiltonian engineering \cite{tyler2023higher,PhysRevLett.131.220803} to investigate the optimal sequences for suppressing decoherence under substantial experimental inhomogeneities. 
We demonstrate the advantages of substituting standard rectangular pulses with shaped, amplitude, and/or frequency modulated pulses. Specifically, we show that the application of chirped pulses that perform rapid adiabatic passage (RAPs) \cite{genov2020efficient} significantly extends $T_2$ in case of NV-ensembles, especially in combination with robust DD sequences. 

We propose and implement NV-ensemble magnetometry experiments using RAPs designed and optimized to be robust, e.g., against large detuning and pulse-amplitude errors. RAPs perform efficient and robust population inversion in the presence of such errors \cite{vitanov2001laser,tannus1997adiabatic,garwood1991symmetric} and thus help prolong the coherence time. Although AC sensing using RAPs in NV-ensembles has been reported in wide-field setups at a high bias magnetic field \cite{genov2020efficient}, our experiments utilize RAP-based DD sensing at a low-bias field regime in a mesoscopic setup. This approach is particularly relevant for emerging applications in nanoscale zero-to-ultra-low-field (ZULF) NMR \cite{barskiy2025zero}, where maintaining high sensitivity at low frequencies is essential despite significant control-field gradients.

\section{Theory}
\label{sec:theory}

\begin{figure}[t]
    \centering
    \includegraphics[width=\columnwidth]{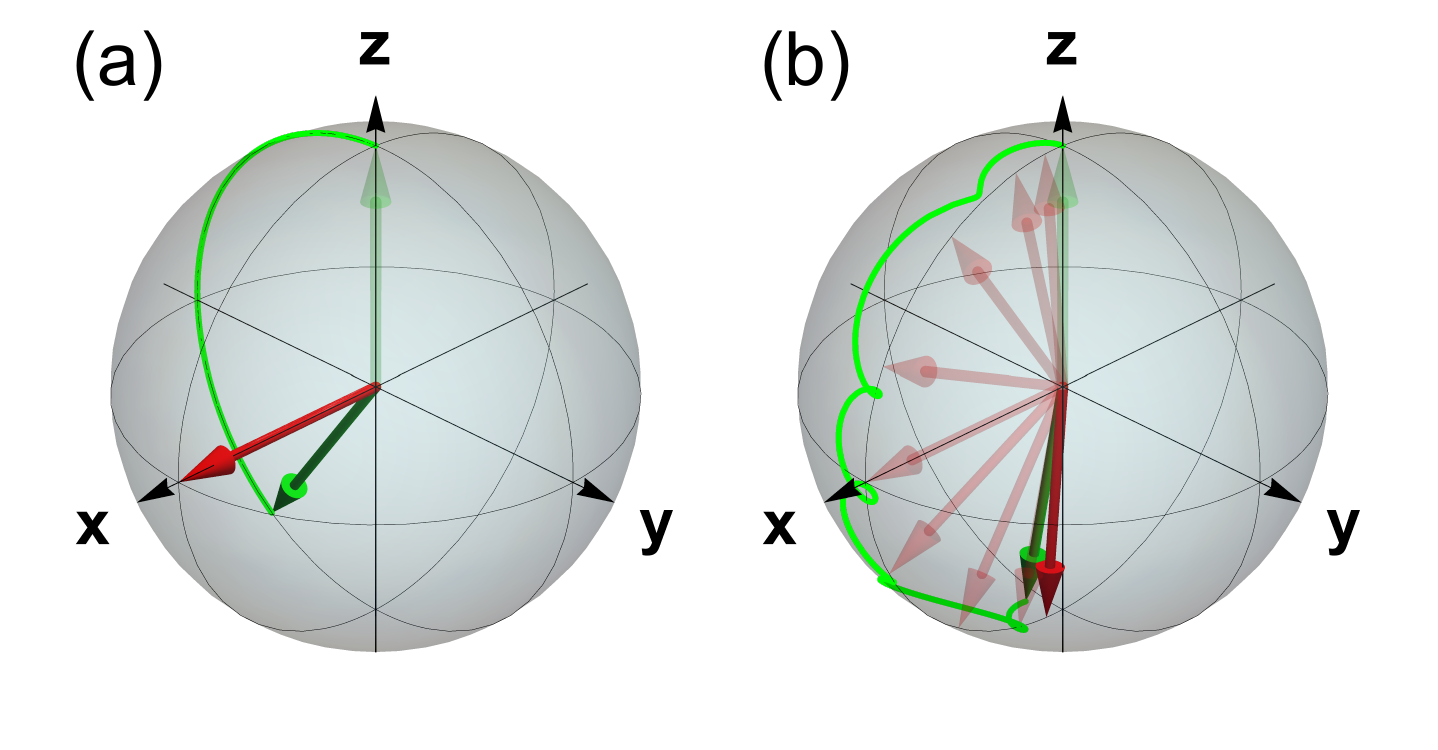}
\caption{Example of the evolution on the Bloch sphere of (a) a rectangular and (b) a RAP pulse in the presence of a 20\% amplitude error, i.e., $\widetilde{\Omega}=\Omega(1+\epsilon)$, $\epsilon=-0.2$, and no detuning errors $\delta_\epsilon=0$. The green semi-transparent (solid) arrow shows the Bloch vector at the beginning (end) of the respective pulse. The red transparent (solid) arrow shows the normalized torque vector during (at the end) of the pulse. The duration of the rectangular pulse is $T_{\text{pulse}}=\pi/\Omega$. The example RAP pulse follows the Allen-Eberly model \cite{Allen-Eberly1987,genov2020efficient}, where $\Omega(t)=\Omega_0 \sech{\left(t/T\right)}$, $\delta(t)=(R/2)\tanh{\left(t/T\right)}$ with peak Rabi frequency $\Omega_0=(2\pi)\,11.45$ MHz, maximum chirp range $R=(2\pi)\,7.5$ MHz, pulse duration $T_{\text{pulse}}=1\,\mu$s, truncation ratio $T_{\text{pulse}}/T=8$, similarly to the RAP pulse in the experimental implementation. It is evident that the final Bloch vector is much closer to the south pole with RAP, demonstrating its robustness for spin inversion. The circles of the RAP pulse path on the Bloch sphere are due to imperfect adiabaticity. }
    \label{fig:rect_rap_bloch_example}
\end{figure}

We consider a two-level system whose evolution is characterized by the Hamiltonian  ($\hbar=1$)
\begin{align}
    H(t) = \frac{\widetilde{\omega}_0}{2}\sigma_z &+ \widetilde{\Omega}(t)\cos{(\omega_0 t+A(t)+\phi)}\sigma_x \notag\\
    &+ g \cos(\omega_{\text{s}}t + \xi)\sigma_z,
    \label{eq:1}
\end{align}
where $\widetilde{\omega}_0=\omega_0+\delta_\epsilon$ is the transition angular frequency of the spin qubit, e.g., consisting of states $|0\rangle$ and $|-1\rangle$ of the electron spin of an NV center (see Fig. \ref{fig:1}a and \cite{JelezkoPRL2004}). It can vary due to an unwanted, typically unknown detuning $\delta_\epsilon$ from its expected value of $\omega_0$. In addition, $\widetilde{\Omega}(t)=\Omega(t)(1+\epsilon)=\mu\widetilde{B}_1(t)$ is the Rabi frequency of the driving field, $\mu$ is the magnetic moment of the spin transition, and $\widetilde{B}_1(t)$ is the actual microwave field amplitude driving the transition, with $\epsilon$ an error term, e.g., due to $\widetilde{B}_1$ inhomogeneity, $\omega_0$ is its carrier frequency, $A(t)=\int_0^t \delta(t^\prime)dt^\prime$ is a frequency modulation of the applied waveform, which depends on the time dependent detuning $\delta(t)$, while $\phi$ is the phase of the driving field. Our goal is to sense the characteristics of a target field with amplitude $g$, carrier angular frequency $\omega_s$, and phase $\xi$. The Hamiltonian in the rotating frame at $\omega_0+\delta(t)$ after applying the rotating-wave approximation takes the form  
\begin{align}
    H_{\text{s}}(t) &= S(t) H(t) S^\dagger(t)-i S(t)\partial_t S^\dagger(t) \notag\\
    &=-\frac{\widetilde{\delta}(t)}{2}\sigma_z + \frac{\widetilde{\Omega}(t)}{2}\left(\cos{(\phi)}\sigma_x+\sin{(\phi)}\sigma_y\right) \notag\\
    &~~~~~~~~~~~~~~~+ g\sigma_z \cos(\omega_{\text{s}}t + \xi),
    \label{eq:2}
\end{align}
where $\widetilde{\delta}(t) \equiv \delta(t) - \delta_{\epsilon}$ with $S(t)=\exp(i \int_0^t H_r(t^\prime)dt^\prime )$, $H_r(t)=\frac{\omega_0+\delta(t)}{2}\sigma_z$. The evolution of the density matrix is then governed by the von Neumann equation $i\partial_t \rho(t)=[H_s(t),\rho(t)]$, where $\rho(t)$ is the density matrix of the system. 
The time evolution can also be visualized by the evolution of the Bloch vector $\mathbf{b}(t)=[2\text{Re}(\rho_{01}),-2\text{Im}(\rho_{01}),\rho_{00}-\rho_{11}]^{T}$ on the Bloch sphere (see Fig. \ref{fig:rect_rap_bloch_example}). It is characterized by 
\begin{align}
    \partial_t \mathbf{b}(t)=\mathbf{\widetilde{\Omega}}_{\text{eff}}(t) \times \mathbf{b}(t),
    \label{eq:bloch_equation}
\end{align}
by a rotation of the Bloch vector around a torque vector $\mathbf{\widetilde{\Omega}}_{\text{eff}}(t)=\frac{1}{2}[\widetilde{\Omega}(t)\cos{(\phi)},\widetilde{\Omega}(t)\sin{(\phi)},-\widetilde{\delta}(t)]^{T}$, with the effective Rabi frequency given by $\widetilde{\Omega}_{\text{eff}}(t)=|\mathbf{\widetilde{\Omega}}_{\text{eff}}(t)|=\sqrt{\widetilde{\Omega}(t)^2+\widetilde{\delta}^2}$. For example, a simple Rabi drive without amplitude errors, no detuning and frequency modulation, and $\phi=0$ would lead to a rotation of the Bloch vector around the $x$ axis of the Bloch sphere with $\mathbf{\widetilde{\Omega}}_{\text{eff}}(t)=\frac{1}{2}[\Omega(t),0,0]^{T}$. Alternatively, $\phi=\pi/2$ would lead to a rotation around the $y$ axis with $\mathbf{\widetilde{\Omega}}_{\text{eff}}(t)=\frac{1}{2}[0,\Omega(t),0]^{T}$ in the absence of errors. Finally, when $\delta(t)\gg|\Omega(t)|$, the $\mathbf{\widetilde{\Omega}}_{\text{eff}}(t)$ is oriented along the $z$ axis of the Bloch sphere, resulting mainly in rotations in the $xy$-plane of the Bloch sphere. The amplitude and detuning inhomogeneities in spin ensembles can lead to significant variation in both the magnitude and orientation of $\mathbf{\widetilde{\Omega}}_{\text{eff}}(t)$, leading to a loss of coherence.

In the absence of a driving field, $\delta_\epsilon$ is typically the main reason for decoherence. It is characterized by the dephasing time $T_2^{\ast}$, which is usually measured by a Ramsey measurement \cite{degen2017}. The distribution of $\delta_\epsilon$ can usually be approximated as Gaussian with a standard deviation $\sigma_{\delta_\epsilon}\approx \sqrt{2}/T_2^{\ast}$ \cite{Gillespie1996,Pascual-WinterPRB2012,senkalla2024germanium,grimm2025coherent,Lim2025optimal}. In our particular experiment with NV centers at ZULF, $\delta_\epsilon$ has additional contributions due to the non-polarized nitrogen nuclear spin, which could lead to shifts of the $\delta_{\epsilon}$ by $\pm (2\pi)\,2.14$ MHz, depending on the state of the nitrogen nuclear spin. These changes and the different orientation of the NV centers lead to a variation of $\delta_\epsilon$ within a range of several MHz, as demonstrated by the ODMR spectrum in  Fig. \ref{fig:6} in the Appendix. 

In order to prolong the coherence time, typical DD experiments include a sequence of rectangular, ideally instantaneous, $\pi$ pulses for dynamical decoupling \cite{viola_dynamical_1999}, i.e., $\Omega(t)=\text{const}$ during the pulse, and $A(t)=0$. These refocus the effect of system-environment interactions, e.g., due to $\delta_\epsilon$, and increase the coherence time to $T_2\gg T_2^{\ast}$ \cite{degen2017}.  When the time separation between the centers of the pulses matches half the period of the carrier frequency of the target field $\tau=1/(2f_s)$, where $f_s=\omega_s/(2\pi)$, the target AC signal is modulated in the interaction frame of the pulses, the sensor spin accumulates a phase due to the target field, and we observe a dip of the signal \cite{degen2017}. This allows us to extract information about the target field, e.g., about its amplitude and frequency \cite{degen2017,genov2020efficient}.

DD usually assumes ideal $\pi$ pulses with $\delta_{\epsilon} \ll \Omega(t)$ and $\epsilon=0$, applied for a duration such that its pulse area is $\int_0^t\Omega(t^\prime)\,dt^\prime = \pi$. In practice, both $\delta_{\epsilon}$ and $\epsilon$ can be substantial and vary significantly in spin ensembles, typically resulting in imperfect fidelity. Figure \ref{fig:rect_rap_bloch_example}(a) shows an example of an imperfect $\pi$ pulse with an amplitude error of 20\%. It is evident that, instead of pointing straight up, the Bloch vector at the end of the evolution has a significant deviation, demonstrating the imperfect fidelity. Applying a high number of imperfect $\pi$ pulses quickly leads to loss of coherence. One can estimate the magnitude of the amplitude errors in a spin ensemble with a Rabi oscillations experiment, where we initially prepare the system in state $|0\rangle$ and drive it with a strong field, which is ideally resonant with the spin transition frequency, i.e., $\Omega(t)\gg\delta_\epsilon$. Figure \ref{fig:1}e shows that the Rabi oscillations decay after only three full periods, implying that the amplitude errors are of the range of $\approx 1/3$. Specifically, the decay due to amplitude inhomogeneity takes the form $\sim \exp\{-(\Omega t\sigma_{\epsilon})^2/2\}$ \cite{Kitamura2026praaplied}, where $ \sigma_{\epsilon }\approx 0.27\pm 0.06$ is the standard deviation of $\epsilon$, which we assume to have a normal distribution \cite{de_raedt_quantum_2012}. 

In this work, we investigate several approaches for improving the coherence time and robustness of DD with respect to errors. 
First, we combine standard rectangular pulses with careful control of the relative shifts in $\phi$ of the individual pulses. These are typically used as control parameters for the design of robust composite pulses and DD sequences  \cite{levitt_composite_1986,wimperis1994broadband,Torosov2011smooth,TorosovPRL2011,genov2014correction,souza2011robust,Souza2012,genov2017arbitrarily,Ezzell23}. An example is the widely used XY8 sequence, where the phases $\phi$ take the values $(0,1,0,1,1,0,1,0)\pi/2$ for each of the eight pulses in the sequence \cite{Maudsley1986xy4,Gullion1990xy,farfurnik2015optimizing,senkalla2024germanium}. 

Second, we replace the standard rectangular pulses with smoothly shaped sine and parabolic pulses, where $\Omega(t)$ is modulated to have the respective shape. These have been shown both theoretically and experimentally to reduce leakage and crosstalk errors due to reduced power-broadening and suppressed off-resonant excitations \cite{BoradjievPRA2013,MihovPRA2024,JoasPRX2025}. The effects of the latter can be significant at low fields and in the case of spectrally dense transitions. %, where Rabi frequency is modulated by a smooth envelope function $E(t)$, $\Omega(t)=\Omega_0 E(t)$ with $E(0)=E(T_{\text{pulse}})=0$. 
%This is motivated by the reduced power broadening achievable with smooth, finite-duration pulse envelopes compared to the standard pulses\cite{BoradjievPRA2013,MihovPRA2024}. 
This is particularly relevant at our operating bias magnetic field of $\approx 1.1$~mT, where closely spaced spin levels of the NV centers in the ensemble make suppressing off-resonant leakage especially beneficial \cite{JoasPRX2025}. Specifically, 
we implemented DD sequences using $sin^n$- and parabolic-power-shaped (order $p$) microwave pulses in combination with the same robust sequences, as for the standard pulses. Full envelope definitions, area-normalization conditions, and associated power-broadening scaling are given in Appendix~\ref{Sec:App_B}, Sec.~\ref{Subsec:Pulse_shaping}. 

Finally, we also test rapid adiabatic passage (RAP) pulses, which have also been demonstrated to be very robust especially against pulse errors \cite{vitanov2001laser,tannus1997adiabatic,garwood1991symmetric,Torosov2011smooth,TorosovPRL2011,MiethPRA2012,Pascual-WinterPRB2012,genov2014correction,genov2020efficient}. Apart from amplitude modulation, these pulses also include frequency modulation $A(t)$, which satisfies the adiabatic condition and ensures robustness in the presence of large amplitude and frequency inhomogeneity errors due to adiabatic evolution. Specifically, in the case of RAP, at the start of the pulse, the drive is far off-resonant, i.e., the detuning $|\delta(t)|\gg\Omega(t)$ dominates, thus the instantaneous torque vector $\mathbf{\widetilde{\Omega}}_{\text{eff}}(t)$ points towards approximately the $\pm z$ direction on the Bloch sphere, as shown in Fig. \ref{fig:rect_rap_bloch_example}(b). As the frequency sweeps through the resonance, the longitudinal detuning term vanishes. Meanwhile, the ramped-up $B_1$ drive tilts the effective field away from the $z$ axis (the torque vector crosses the $xy$-plane). If the sweep is slow enough, the Bloch vector follows $\mathbf{\widetilde{\Omega}}_{\text{eff}}(t)$ continuously, as shown in Fig. \ref{fig:rect_rap_bloch_example}(b). Therefore, at the end of the pulse, the sweep places $\mathbf{\widetilde{\Omega}}_{\text{eff}}(t)$ on the opposite pole of the Bloch sphere, leading to an adiabatic inversion of the population of the two states.

The condition for adiabatic evolution is typically $\Omega^{2} T_{\text{pulse}}/ R \gg 1$  \cite{vitanov2001laser,doll2013adiabatic,genov2020efficient}, where $\Omega$ is the Rabi frequency, $T_{\text{pulse}}$ is the pulse duration, and  $R$ is the chirp range. We note that in our work we apply RAP pulses that follow the Allen-Eberly model due to its preferable adiabaticity with $\Omega(t)=\Omega_0 \sech{\left[(t-t_{c,k})/T\right]}$, $\delta(t)=(R/2)\tanh{\left[(t-t_{c,k})/T\right]}$, where $t_{c,k}$ is the k-th pulse center, and $T$ is a characteristic time, so the adiabatic condition becomes $\Omega_0^{2} T/ R \gg 1$ \cite{Hioe1984solution,Allen-Eberly1987,Kyoseva2006analytic,genov2020efficient}. A detailed discussion on the RAP mechanism and the adiabatic condition for different RAP pulses can be found in \cite{genov2020efficient}. We note that RAP can perform reasonably well even in the case of imperfect adiabaticity \cite{Benseny2021adiabatic_theorem,Genov2023adiabatic} (see also Fig. \ref{fig:rect_rap_bloch_example}). Similarly to the DD protocols with standard and smooth pulses above, we also combine the RAP pulses with robust DD sequences, using the phases $\phi$ as control parameters to improve the robustness even further \cite{TorosovPRL2011,genov2014correction,genov2017arbitrarily,genov2020efficient}. As the pulse imperfections, including non-adiabaticity, are then compensated by the particular choice of the pulses' phases, we optimize the RAP pulse parameters experimentally rather than relying on the usually too stringent adiabatic condition alone \cite{genov2020efficient,Benseny2021adiabatic_theorem}.

\section{Experimental Demonstration}
\label{sec:Experiment}

We conduct our experiments with an ensemble of NV centers
in diamond. The measurements are performed in 
a custom-built wide-field setup, comprising an optical excitation path and a comprehensive data acquisition system at a bias magnetic field of 1.1 mT. Higher magnetic fields are challenging to achieve due to technical limitations in our setup, which is originally designed for vector magnetometry. The system and experimental
setup are described in Fig.\ref{fig:1}(a)-(c). Our qubit consists of states $|0\rangle$ and $|\text{-1}\rangle$ of the ground state of the electron spin of the NV centers in the ensemble, which we drive resonantly with a microwave field. We note that the electron ground spin state $|\text{+1}\rangle$ is detuned by only $\Delta/(2\pi)\approx 61$ MHz at this particular magnetic field, which might lead to leakage errors, i.e., unwanted population transfer to $|\text{+1}\rangle$ due to off-resonant excitation.  

In the optical path, we used a green laser beam of $532$~nm wavelength from Coherent (Verdi V6) to excite the NV centers in the ensemble. A detailed schematic of the full optical setup is provided in Appendix~\ref{sec:experimental_details} (Fig.~\ref{fig:scheme}). To begin with, the laser beam is focused onto an Acousto-Optic Modulator (AOM) (Gooch \& Housego 3250-220).  
After the AOM, the diffracted light is recollimated, and an iris is used to spatially select the first-order diffraction beam while blocking all other diffraction orders. Then, we use two plano-convex lenses in a telescopic configuration to expand the beam waist. Finally, the magnified beam focuses onto the sample, which contains an ensemble of NV centers (HPHT $99.995\%$ $^{12}$C- enriched, with $(111)$ orientation), and the spot diameter was $47\,\mu\text{m}$. The NV and the $\text{P}1$ concentrations, measured via EPR, are $0.18$~ppm and $1.2$~ppm, respectively.
The detection region is housed within an optical cage assembly, as illustrated in Fig.~\ref{fig:1}-(c). The sample (in pink) is placed on a compound parabolic condenser (CPC in light blue), to improve collection efficiency~\cite{PhysRevX.5.041001}. 
%The cage also supports the home-grown resonator for the $B_1$ drive that is placed close to the surface of the sample.
The resulting fluorescence (red arrow) is directed through a long-pass (LP) filter positioned between two spherical lenses (SL) and is subsequently detected by the photodiode (PD10A2 Thorlabs, denoted as PD) at the end of the cage. The PD carries the fluorescence signal to the computer. 

This entire assembly is placed inside the homogeneous region of a 3-axis Helmholtz coil (Bartington HC16) for the bias field. We apply a static bias magnetic field with a magnitude of $1.1$ mT, which has equal contributions along the $x,y,$ and $z$ axes (represented by yellow, purple, and blue rings, respectively). Microwave delivery is achieved by a custom-designed resonator (PCB), shown in Fig.~\ref{fig:1}-(b), which is based on some modifications of the Lenz-lens structure~\cite{sasaki2016broadband} (see Sec.~\ref{sec:antenna} in the Appendix for design specifics). Then comes the data acquisition framework, where the analog signal from the photodiode is connected to the computer and converted to digital via a high-speed ADC (Spectrum Instrumentation M4i.4420-x8). Microwave generation, pulsing, and synchronization control are performed with an arbitrary waveform generator (AWG Tektronix AWG70001A-16 Gs). The entire measurement process and data acquisition was performed with the Qudi measurement suite~\cite{binder2017qudi}.

\paragraph{Initial Characterization.---} The Rabi frequency was set at $\Omega(t)/(2\pi)\approx \,11$ MHz, with a carrier frequency tuned to the resonance dip at $\omega_0/(2\pi)\approx\,2.839$ GHz, identified via ODMR (see Fig. \ref{fig:6} in Appendix\ref{sec:odmr_details}). The dephasing time $T^{*}_{2}$ from the Ramsey measurement is determined to be $1.68 \pm 0.27$ $\mu s$. We note that homogeneity of the bias magnetic field improved when we used the Helmholtz coil \cite{beaudoin2013enhanced}, yielding a factor of $\sim 8.5$ increase in $T^{*}_{2}$ compared to measurements using permanent magnets. The $B_1$ drive was amplified, using the amplifier AR 75S1G6C at $50\%$ gain. The Rabi oscillation can be seen in Fig.~\ref{fig:1}(e). 
The Rabi contrast was $17.84\%$, and all subsequent measurements were normalized to this value. As noted in the previous section, we observe a fast decay of the Rabi oscillations due to significant amplitude inhomogeneity with a standard deviation $ \sigma_{\epsilon }\approx 0.27\pm 0.06$. The $B_{1}$ field inhomogeneity can in principle be substantially mitigated using 3-D resonators, which offer superior uniformity compared to planar ones, as we have demonstrated in previous work \cite{rezinkin2024uniform}. However, such systems often have restricted optical access, thus limiting the preparation and readout efficiency. To circumvent this problem, we proceeded with the planar structures and utilized the optimized RAP pulses (RAPs) that are more robust to inhomogeneity. The Rabi contrast improves to $24.5\%$ with RAPs, as demonstrated in Fig.~\ref{fig:1}(f), where we replace the simple constant Rabi driving field, used in Fig.~\ref{fig:1}(e), with a sequence of identical RAPs. The robustness of RAPs is also evident in Fig.~\ref{fig:1}(f), as we can observe more than twenty spin flips with RAP pulses before decay in comparison to only about six with standard Rabi driving. Other fundamental spin properties were measured as $T_1 = 6.2 \pm 0.1$ ms and Hahn echo $T_2 = 15.5 \pm 0.6$ $\mu s$ (see Appendix \ref{Sec:App_B} for details). 
 
\paragraph{Optimization of RAPs.---} RAPs were optimized with a combination of numerical simulations and iterative experimental characterization. First, the RAP pulse parameters were optimized, taking the adiabaticity condition into account, followed by a numerical simulation of the population transfer efficiency. The numerical simulations and the results from the ODMR measurement indicated that a chirp range of $R/(2\pi)\approx7.5$ MHz would cover all three transitions, arising from the non-polarized nitrogen nuclear spin; see Fig. \ref{fig:6} in Appendix \ref{sec:odmr_details}). Numerical simulations were also used to optimize the truncation $T_{\text{pulse}}/T=8$, i.e., the ratio between the pulse duration and the characteristic time of the pulses. It characterizes the smoothness of the pulses at their truncation points with a higher truncation ratio improving adiabaticity at the pulse edges at the expense of longer pulse durations \cite{genov2020efficient}. Then, we performed experimental optimization of the pulse duration $T_{\text{pulse}}$ for $R/(2\pi)=7.5$ MHz by maximizing the contrast of the Rabi oscillations in Fig. \ref{fig:1}(f). The experiments showed stable performance for pulse durations between 0.5 and 2 $\mu$s (see Fig. \ref{fig:10}(a) in the Appendix) with optimal duration at $T_{\text{pulse}}\approx 1\,\mu$s. Then, we fixed $T_{\text{pulse}}=1\,\mu$s and optimized experimentally the chirp range R, using the truncation ratio $T_{\text{pulse}}/T=8$. We achieved maximum Rabi oscillations contrast with $R/(2\pi)= 20$ MHz but the longest coherence time was obtained with $R/(2\pi)= 7.5$ MHz, most likely due to leakage when applying many pulses with the larger chirp range (see Fig. \ref{fig:10}(b-d) in the Appendix). The optimal sensitivity for quantum sensing typically requires maximization of $T_2$, so we chose $R/(2\pi)=7.5$ MHz for subsequent experiments. The details of the experimental optimization are described in Appendix~\ref{sec:opt_raps}.

\paragraph{Dynamical decoupling optimization for quantum memory.---} Next, we compared the performance of the standard rectangular, smooth and rapid adiabatic pulses in combination with robust DD sequences \cite{Gullion1990xy,souza2011robust,genov2017arbitrarily}. The goal was to find the optimal sequence in terms of quantum memory time and thus sensitivity for quantum sensing. The time between the centers of the pulses $\tau$ was optimized with the XY8 sequence to find which pulse separation gives the longest memory time when repeating the sequence multiple times. In the case of perfect instantaneous $\pi$ pulses, the pulse separation should typically be as small as possible \cite{Pascual-WinterPRB2012,senkalla2024germanium,grimm2025coherent,Lim2025optimal}. However, in the presence of pulse errors and inhomogeneities, a higher number of pulses usually leads to error accumulation and loss of fidelity, so longer inter-pulse free evolution times become beneficial \cite{genov2017arbitrarily}. Thus, there is an optimal inter-pulse delay, taking into account the trade-off between the two error mechanisms \cite{genov2017arbitrarily}. 

\begin{figure}[t]
    \centering
    \includegraphics[width=\columnwidth]{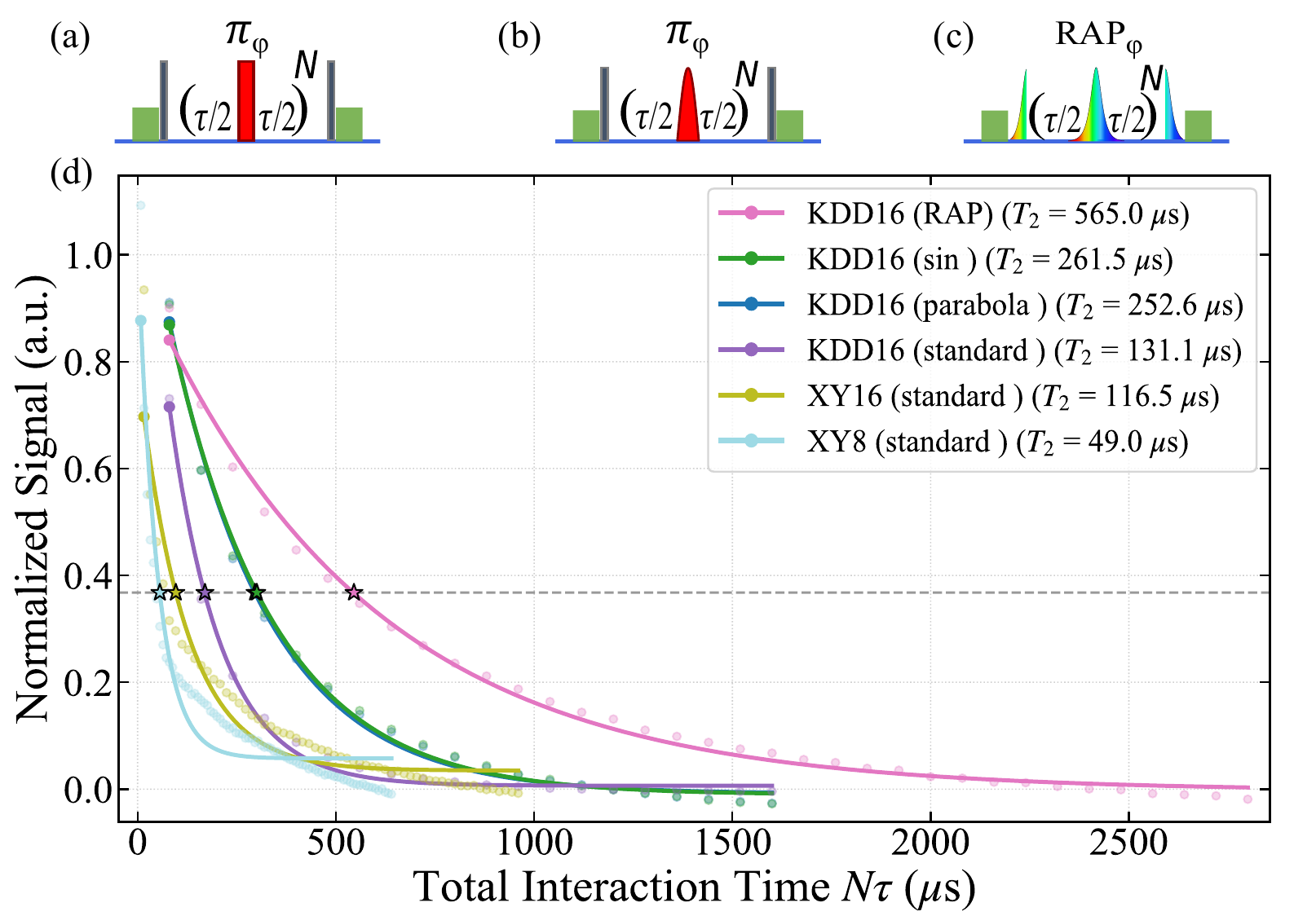}
    \caption{Schematic illustrations of the applied pulses in the respective sequences for (a) standard rectangular pulses, (b) sine- and parabola-shaped $\pi$ pulses, and (c) RAP pulses. (d) Signal decay curves for various dynamical decoupling (DD) protocols, comparing XY8, XY16, and KDD16 sequences implemented with standard, parabola-shaped, and sine-shaped pulses. The horizontal axis represents the total interaction time $t=N\tau$, where we vary the sequence order $N$, i.e, the number of times a sequence is repeated, and the vertical axis displays the photoluminescence signal normalized to the contrast of standard Rabi oscillations. The time separation between the centers of the pulses $\tau$ was optimized using the standard XY8 sequence and fixed at $1\,\mu s$. We fit the data with the single exponential decay function as $S(t)=a \, e^{-t/T_2} + c$ where $a$ is the initial amplitude and $c$ is a constant offset. The decay constant $T_2$ represents the coherence memory time. The star markers on each curve indicate the $1/e$ decay points, facilitating a direct visual comparison. The KDD16-RAP sequence exhibits the longest memory time, yielding a $T_2$ of $565$ $\mu s$. 
    }
    \label{fig:2}
\end{figure}

We experimentally found the optimal value to be $\tau \approx1\mu s$ and used it for subsequent memory time measurements, where we varied the number of repetitions of each DD sequence, i.e., the total number of applied $\pi$ pulses, keeping all other parameters fixed. First, we applied the widely used XY8 sequence with standard rectangular pulses, obtaining a memory time $T_2\approx 49\,\mu$s (see Fig. \ref{fig:2}). We note that  the limit for the memory time for $\tau=1\,\mu$s is set by the Carr-Purcell-Meiboom-Gill (CPMG) sequence at $T_{1,\rho}\approx 902\,\mu$s \cite{carr1954effects,MeiboomGill1960}. CPMG is robust to pulse errors when the initial state is aligned with the axis of rotation of the $\pi$ pulses but otherwise it loses fidelity rapidly \cite{souza2011robust}. Therefore, while it could be applied for frequency sensing, it cannot be used for storing arbitrary states, i.e., for a quantum memory. It is also not efficient for amplitude sensing, where the state of the system changes during the DD sequence, as the qubit accumulates a phase, which depends on the AC field amplitude \cite{pham2012enhanced,senkalla2024germanium,genov2017arbitrarily}.

The main limitation to the performance of XY8 was pulse errors due to inhomogeneities, as we observed a significant improvement when we modified the pulses' phases with the ones of the more robust XY16 \cite{Gullion1990xy} or the KDD16 sequence \cite{souza2011robust}, i.e., $T_2\approx 117\,\mu$s and $131\,\mu$s, respectively. KDD16 is based on a composite pulse, first proposed by Tycko and Pines \cite{Tycko1986iterative} and later generalized as the U5b pulse in \cite{genov2014correction}. As proposed by Knill \cite{Souza2012}, the constituent pulses of the composite pulse are time separated. We replace each $\pi$ pulse in XY16 with this time separated composite pulse to obtain the KDD16 sequence \cite{souza2011robust}. We obtained a similar improvement in memory times with the UR sequences \cite{genov2017arbitrarily} and when using the U5a pulse from \cite{genov2014correction}, embedded in XY16, although KDD16 slightly outperformed, most likely due to the specific experimental errors in our setup. 
As a next step, we also tried to improve performance by replacing the individual rectangular pulses in the DD sequences with composite pulses, which contained pulses that were not time-separated, e.g.,  BB1 \cite{wimperis1994broadband}, U5a, and U5b \cite{genov2014correction}, embedded in XY8 or XY16. Overall, we found that the DD sequences with time-separated composite pulses perform better than the ones with non-time-separated composite pulses.  

Next, we applied DROID (Disorder RObust Interaction Decoupling) and DIRAC2 (DIsorder Robust AC with period Two) sequences \cite{tyler2023higher,PhysRevLett.131.220803} that compensate both spin-environment and spin-spin interactions, which might be expected in a dense ensemble of NV centers. These are constructed using higher-order Hamiltonian engineering, which extends average Hamiltonian theory to derive explicit cancellation rules for higher-order terms in the Magnus expansion~\cite{ulrich1976high}. Both DROID and DIRAC2 significantly underperformed in terms of memory time (see Fig.~\ref{fig:11} in the Appendix) in comparison to KDD16 and the other robust sequences, which are designed to compensate disorder and spin-environment interactions only. Thus, we can conclude that NV spin-spin interactions are not the main source of decoherence in our specific experiment in comparison to inhomogeneities and other error sources.

\begin{figure}[t]
    \centering
    \includegraphics[width=\columnwidth]{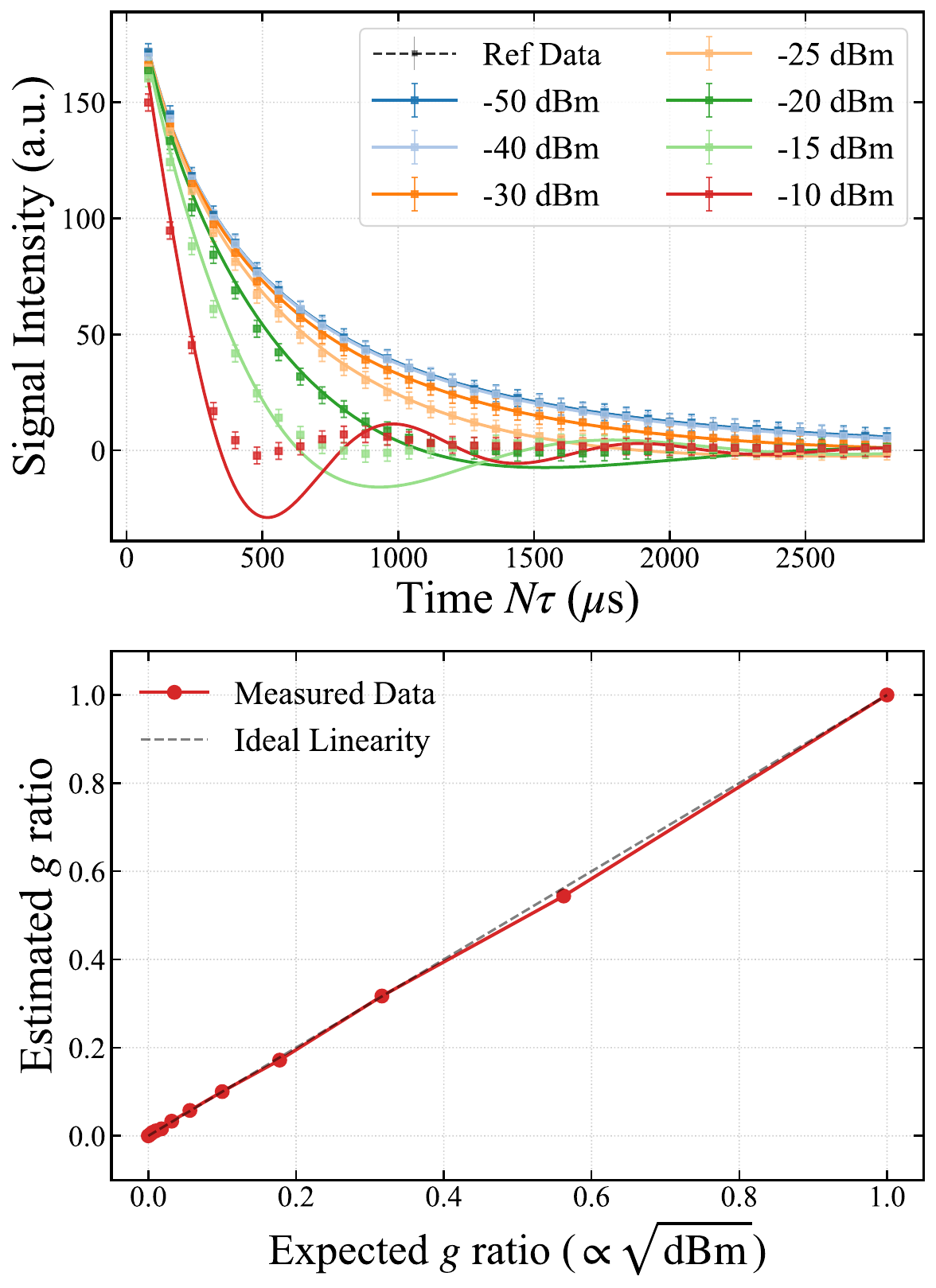}
    \caption{Comparison of the signal intensity decay with KDD16-RAP sequence in the presence of an AC magnetic field with varying amplitude. (Top) Coherence decay curves in the presence of a $500$~kHz test signal applied at different power amplification, from $-50$ to $+10$ dBm. The \enquote{Ref} data (dotted grey) represent the reference signal in the absence of an externally applied field. The filter function was set at $\tau = 1 \mu $s to perform order-N sweeps. The extended coherence time $T_2$ and enhanced contrast $C$, provided by the RAPs, lead to significant improvement in sensitivity, according to the relation $\eta \propto 1/(\gamma C \sqrt {T_2})$, where $\gamma$ is the gyromagnetic ratio. The decay curves are fitted with a stretched exponential decay modulated by a zeroth-order Bessel function of the first kind $J_0(2\pi gt)$, following Eq.~\eqref{eqtn2}. %, where the \enquote{g} parameter is the coupling strength through which the signal is coupled to the external field. 
    (Bottom) The signal amplitudes $g$ were estimated by normalizing them to that of the estimate, obtained at $0$ dBm amplitude amplification ($g/g_{\text{0 dBm}}$). The resulting ratios were plotted against the corresponding expected amplitude ratios of the applied test fields. Further details are provided in Appendix~\ref{sec:amplitude_details}.}
    \label{fig:3}
\end{figure}

As a next step, we replaced the standard rectangular pulses with smooth amplitude modulated $sin^n$ and parabola pulses in order to reduce leakage and crosstalk errors. These are often present in low magnetic field experiments \cite{JoasPRX2025}, as also noted in the previous section. The memory time was improved by a factor of two, to $262$ $\mu$s with $sin^1$-shaped pulses and KDD16 in comparison to the combination of the standard rectangular pulses and KDD16 (see Fig. \ref{fig:2}). The parabola pulses ($p=1$, see Appendix \ref{Subsec:Pulse_shaping}) demonstrated similar improvement to  $251\,\mu$s although it was slightly lower than the one obtained with the $sin^1$ pulses. Henceforth, we show only the results of KDD16 as the relative performance of the other robust sequences with $sin$ and parabola-shaped pulses was usually similar to the case with the rectangular pulses. 
We also tested $sin^n$ (n=2,3) and parabola-shaped (order-$p=2,3$) pulses but they demonstrated suboptimal performance, most likely due to their longer duration and smaller bandwidth.   

Finally, we replaced the amplitude modulated pulses with RAPs, which are combined with the robust DD sequences. The experimental results showed that KDD16-RAP yielded an increase of the memory time to $565$ $\mu$s, as shown in Fig. \ref{fig:2}. This is more than double the memory time, compared to KDD16 using $sin^1$-shaped pulses, and more than an order of magnitude enhancement in comparison to the widely used XY8 with standard, rectangular pulses. The combination of other robust DD sequences \cite{souza2011robust,genov2017arbitrarily} with RAP pulses led to a comparable improvement in coherence time with KDD16-RAP performing slightly better than the other robust DD-RAP sequences in our experimental setup in terms of memory time. 

\paragraph{Quantum sensing of AC-field amplitude.---} Next, we evaluate the performance of the robust DD sequences for quantum sensing of an external AC target field. The latter was delivered near the sample via a custom-built RF resonator based on the Omega structure \cite{bayat2014efficient}. The frequency $f_{s}$ of the external target signal to be sensed, generated by a signal generator (Rohde \& Schwarz SMIQ03B), was set to $0.5$ MHz and we varied its amplitude from $-50$ to $10$ dBm.  This frequency is the same as the one denoted in Hamiltonian~\eqref{eq:1} and related via $\omega_s=(2\pi) f_{s}$. Initial sensing was characterized using an XY8 order-N scan, where the inter-pulse delay was set to $\tau = 1/(2f_{s}) = 1 \mu s$. This configuration serves as a narrowband filter function tuned to the $f_{s}$. Subsequently, the KDD16-RAP was employed to sense the same signal, with $T_{\text{pulse}}=1/(2f_{s})=1 \mu$s \cite{genov2020efficient}.
For both cases, the resulting sensor response was modeled using a stretched exponential decay modified by a zeroth-order Bessel function, $J_0(2\pi gt)$~\cite{Laraoui2010,Spohn2025mcs}, to account for coherent phase accumulation according to Equation~\eqref{eqtn2} in Appendix~\ref{sec:amplitude_details}.

We observe clearly that the changes in the decay rate due to tiny changes in the AC field amplitude can be detected better with KDD16-RAP than with standard XY8. Specifically KDD-RAP can detect a difference from the decay curve from the no-signal case even for a weak test field (-35 dBm, see Appendix, Fig. \ref{fig:13}). This is not possible with the standard rectangular XY8 sequence, where only signal strength of around (-15 dBm) could be distinguished from the decay curve when no signal was applied.

As a next step, we varied the target field amplitude and characterized the correlation between the estimated value of $g$ and the amplifier attenuation setting ($g\propto \sqrt{P}$). Figure \ref{fig:3} (bottom) shows that the estimated values of $g$ follow almost perfectly the ideal linear curve with a slope, which is expected from theory, thus validating the fitting procedure and the applicability of KDD16-RAP to amplitude sensing with high inhomogeneities \cite{degen2017,taylor2008}.

\begin{figure}[t]
    \centering
    \includegraphics[width=\columnwidth]{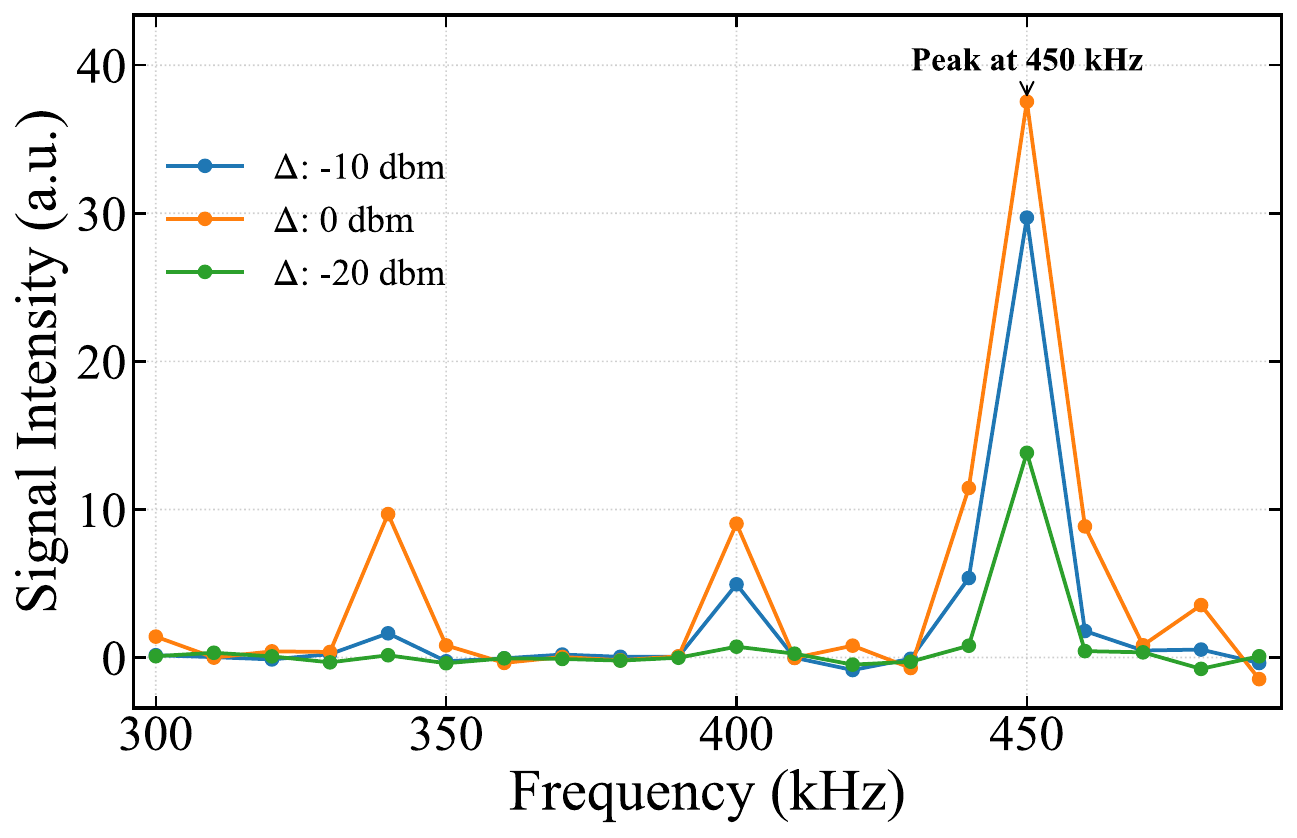}
    \caption{Intensity of the signal difference, obtained with KDD16-RAP sequence, repeated $N=8$ times, in the presence and absence of an external target field with a test signal at $f_s=450$ kHz, measured at three power levels vs. probe frequency $f_p=1/(2\tau)$. The primary resonance at $450$ kHz scales with power, as do the spurious harmonics observed at $340$ kHz and $400$ kHz (see text).}
    \label{fig:4}
\end{figure}

\paragraph{Frequency sensitive measurements.---} 
Apart from amplitude sensing, the optimized DD sequences can also be used for sensing the frequency of the target AC signal. Usually, CPMG is the optimal choice for frequency sensing as its resolution is limited by $T_{1,\rho}$ \cite{degen2017}. Nevertheless, we evaluate the system's frequency response with KDD16-RAP for completeness. To do so, we performed a scan using the DD-RAP protocols in which we varied the time between the centers of the pulses $\tau=T_{\text{pulse}}+\tau_{\text{ip}}$ by changing the interpulse delay between two subsequent RAP pulses $\tau_{\text{ip}}$, while keeping $T_{\text{pulse}}=1\,\mu$s. We note that usually it is optimal to keep $\tau_{\text{ip}}=0$ and vary $T_{\text{pulse}}$ as this improves the adiabaticity of the RAP pulses \cite{genov2020efficient} but the simpler approach where we varied $\tau_{\text{ip}}$ was also sufficient in our case. We define a probe frequency as $f_p=1/(2\tau)$, i.e., this is the frequency where we expect the filter function of our DD sequence to allow the detection of a signal \cite{degen2017}. For each measurement, we conducted two scans of the frequency $f_p$ by varying $\tau_{\text{ip}}$ and thus $\tau$ -- with and without the external signal. The response was quantified using the differential signal from the two measurements. %, defined as $\Delta = Signal_{Test_{on}}-Signal_{Test_{off}}$, where ‘off’ state represents the intrinsic background signal of the system. 
Figure \ref{fig:4} shows the resulting signal response when $f_p$ was varied within the $300$ to $500$~kHz range. We observed the main peak at the expected frequency $f_p=f_s=450$~kHz along with two additional smaller peaks at approximately $\approx 3f_s/4\approx 340$ kHz and $\approx 8f_s/9\approx 400$~kHz (Fig.~\ref{fig:4}), when the signal amplitude was relatively strong, most likely due to spurious harmonics \cite{Loretz2015spurious}. 
To investigate the spurious harmonics response of KDD16-RAP, we applied phase randomization of the different orders of the sequence, i.e., random global phases are added to all phases within each repeated block of the DD, while the relative phases within each block were unchanged~\cite{wang2019randomization}. The measurements were repeated with different random phases and the resulting signal was averaged over these repetitions. Despite the phase randomization the additional harmonics for strong fields were still present in the spectrum for KDD16-RAP, i.e., likely due to the KDD16-RAP filter function as the RAP pulses are non-instantaneous. 

Next, we observed that the magnitude of the peaks increases with the amplitude of the external signal, confirming the sequence's sensitivity to the strength of the applied field, as also shown in the previous section. The precise overlap between the frequency of the applied test signal and the frequency where we observed the main peak demonstrates the high spectral selectivity of the KDD16-RAP sequence. 
 
Consequently, we can conclude that these DD protocols and pulses are applicable to a wide variety of experimental platforms, including those subject to large detuning and amplitude errors.

\section{Conclusion}

In this work, we investigated an extensive number of DD protocols to prolong the memory time and thus improve sensitivity of quantum sensing with spin ensembles with significant amplitude and frequency inhomogeneities. These included applying standard rectangular, amplitude modulated and RAP pulses in combination with state-of-the-art robust DD sequences. The RAP pulses are robust to pulse errors, enabling efficient spin state inversion in the presence of significant inhomogeneities. We showed experimentally that we could improve the memory time in an ensemble of NV centers in diamond by more than an order of magnitude with RAP pulses and robust DD sequences, e.g., KDD16, in comparison to the widely used XY8 sequence with standard, rectangular pulses. 

We also performed amplitude and frequency sensing of non-phase-matched external ac signals using KDD16-RAP sequence, which enhanced significantly the contrast and memory tim in comparison to the standard DD protocols, thus improving sensitivity, e.g., for AC amplitude sensing. 
The protocol is also broadly applicable to other spin-ensemble platforms and to systems affected by significant amplitude and/or frequency inhomogeneity, offering a practical route toward more robust DD and quantum sensing in inhomogeneous environments.  

\begin{acknowledgments}
We thank our collaborators and funding agencies for their support. We also thank Oleg Rezynkin for his valuable input on the construction of the microwave/RF resonating structures. The work was supported by the German Federal Ministry of Research, Technology and Space (BMFTR) via future cluster QSENS and projects EXTRASENS (13N16935), DIAQNOS (No. 13N16463), quNV2.0 (No. 13N16707,  Deutsche Forschungsgemeinschaft (DFG) via projects 386028944, 387073854, 445243414, 491245864, 499424854, 532771161, 546850640, and joint DFG/JST ASPIRE program via project 554644981, European Union’s HORIZON Europe program via projects   SPINUS (No. 101135699), CQuENS (No.101135359), QCIRCLE (No. 101059999) and FLORIN (No. 101086142), QUANTERA via Project MultiQS  and Mf-QDS European Research Council (ERC) via Synergy grant HyperQ (No. 856432), IQST and Carl-Zeiss-Stiftung.

\end{acknowledgments}

\bibliographystyle{apsrev4-2}
\bibliography{references_submit}
% \documentclass[
%  reprint,
%  superscriptaddress,
%  amsmath,amssymb,
%  aps,
%  prapplied,
% ]{revtex4-2}
% \usepackage{graphicx}
% \usepackage{dcolumn}      % Align table columns on decimal point
% \usepackage{bm}
% \usepackage{hyperref}
% \usepackage{caption}
% \usepackage{subcaption}
% \usepackage{physics}
% \graphicspath{{Images/}}
% \usepackage{stfloats}
% \usepackage{placeins} 
% \usepackage{csquotes}
% \usepackage{float}
% \begin{document}

\pagebreak
\appendix
\setcounter{figure}{0}
\renewcommand{\thefigure}{A.\arabic{figure}}
\renewcommand{\theequation}{A.\arabic{equation}}

\section{Experimental details}\label{sec:experimental_details}
\subsection{Experimental setup}
\begin{figure}[t]
    \centering
    \includegraphics[width=\columnwidth]{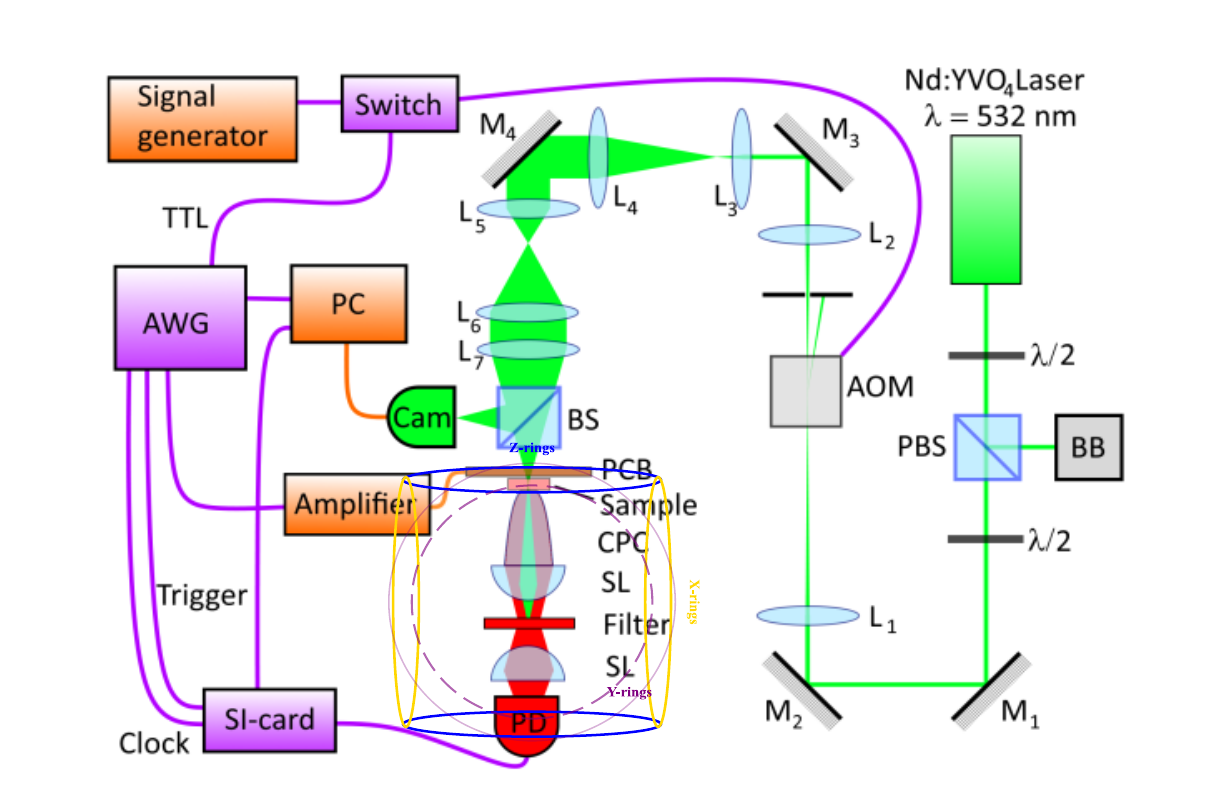}
    \caption{Magnetometer setup:  The AWG is used as a pulse generator for the laser pulses through a microwave switch and for the microwave antenna. It also serves as the reference clock and the
    trigger signal for the data acquisition. Labels: PC: Acquisition computer; TTL: Transistor-transistor logic; Si-card: Analog to digital converter. }
    \label{fig:scheme}
\end{figure}

The schematic of the experimental setup is shown in the Fig.~\ref{fig:scheme}. In the optical part, the green ($532$ nm) beam from the laser Nd:YV$O_4$ (Coherent Verdi V6) passes through a half-wave-plate ($\lambda/2$), then a polarizing beam-splitter (PBS) to select one polarization, with the other polarization mode blocked by a beam blocker (BB). A second $\lambda/2$-plate then returns the beam to its original polarization. The beam is subsequently deflected by two mirrors, $M_1$ and $M_2$, into the next stage, where a lens ($L_1$) focuses it onto an acousto-optic modulator (AOM) with a spot diameter of $100$ $ \mu m$. The AOM is controlled via a signal generator (Siglent SSG3021X). The first-order diffracted beam is selected by blocking the zeroth-order with an iris; the measured rise-time of the AOM is $21.2$ ns. This ﬁrst-order beam is then collimated by the focusing lens ($L_2$) and deflected by mirror ($M_3$) into a telescopic configuration with lenses $L_3$ and $L_4$ to increase the beam-waist. The expanded beam is further deflected by a mirror $M_4$ through to a second telescopic configuration with lenses $L_5$ and $L_6$ for additional beam-waist expansion. Lens $L_7$ then focuses the beam onto the sample; prior to this, the beam-size and shape can be monitored with a camera (CAM) by deﬂecting a portion of the beam via a beam splitter (BS). The focused beam passes through the hole on the PCB before illuminating the sample. The fluorescence from the diamond sample is emitted in all directions, but most of it is reflected back into the diamond due to total internal reflection at angles larger than the critical angle of $24.36^ \circ$, resulting from the diamond's high refractive index, $n_d=2.42$~\cite{weber2002handbook}. To increase the collection efficiency, we use a compound parabolic concentrator (CPC, Edmund optics) to couple light out of the diamond; the sample is glued to the small end of the CPC. The parabolic shape of the concentrator guides the red fluorescence towards the broader end with minimal loss. The collected beam is focused by a spherical lens (SL) and passed through a long pass filter to remove residual green light, before being focused onto the photodiode (PD) via a second spherical lens. 
The acquisition part, described in the main text in Sec.~\ref{sec:Experiment}, comprises the components labeled in Fig.~\ref{fig:1} (c). The coloured rings represent the three-axis Helmholtz coil surrounding the detection region. 
\paragraph{Sample.---} The diamond sample was fabricated by the temperature gradient method of high-pressure and high-temperature (HPHT) technique, followed by electron-beam irradiation and subsequent annealing. Growth took place at $5.5$~GPa and $1350\,^\circ\text{C}$, using an Fe-Co-Ti solvent together with a $^{12}$C-enriched carbon source, prepared by the pyrolysis of $99.999\%$ $^{12}$C-enriched methane. The as-grown crystal was subjected to $2$~MeV electron irradiation at room temperature, followed by vacuum annealing at $1000\,^\circ\text{C}$ for $2$ hours.
Secondary ion mass spectrometry (SIMS) confirmed a $^{12}$C abundance of $99.995\%$. A $(111)$-oriented plate was extracted by laser-cutting and subsequently polished.

\subsection{Optically detected magnetic resonance (ODMR) characterization}\label{sec:odmr_details}
\begin{figure}[t]
    \centering
    \includegraphics[width=\columnwidth]{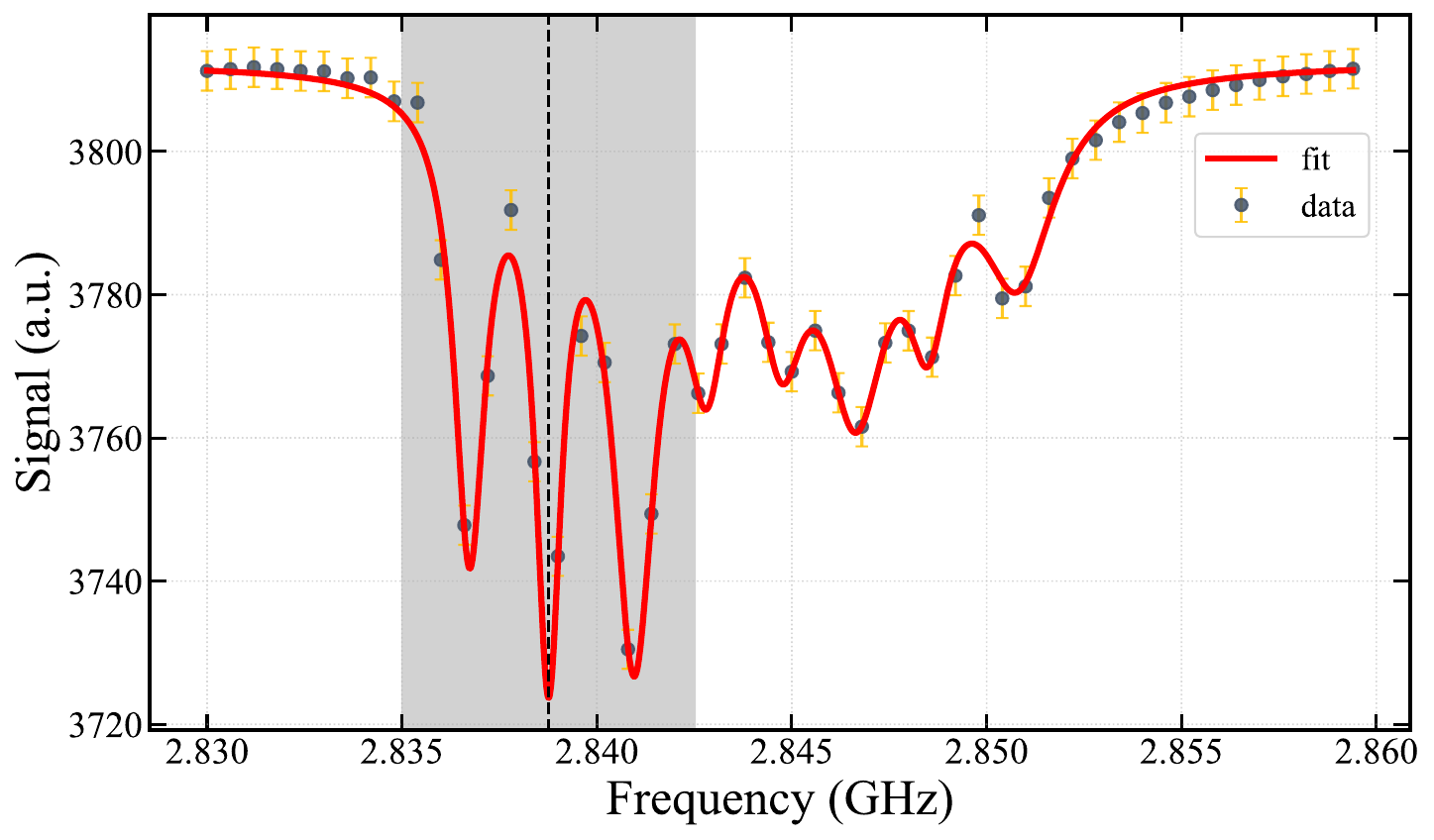}
    \caption{Pulsed ODMR spectrum of the NV ensemble at a bias field of $1.1$~mT. Grey points are the measured fluorescence signal with error bars representing the standard deviation; the red curve is a fit of Lorentzian resonances, with equation $S(f) = S_0 + \sum_{i=1}^{n_R} a_i \, \frac{\sigma_i^2}{(f - f_i)^2 + \sigma_i^2}$, where $S(f)$ is the photoluminescence signal as a function of the microwave frequency $f$, and $S_0$ is the baseline signal. Each of the $n_R$ resonances is described by a center frequency $f_i$, half-width at half-maximum (HWHM) $\sigma_i$, and amplitude $a_i$. The dashed line marks the central hyperfine component at $2.83876$~GHz, and the shaded region ($\text{Range} = 7.5$~MHz wide) spans the $^{14}$N hyperfine triplet of a single NV orientation; the remaining dips arise from the other crystallographic orientations, illustrating the closely spaced resonances susceptible to off-resonant microwave leakage.}
    \label{fig:6}
\end{figure}

To characterize the NV ensemble's spin resonance structure under the applied bias magnetic field, we performed pulsed ODMR spectroscopy. In this measurement, NV population is first initialized into $\ket{m_s=0}$ by a laser pulse, followed by a microwave pulse of fixed duration whose frequency is stepped across a range. Compared to a continuous-wave ODMR the pulsed scheme avoids power-broadening, yielding narrower resonance linewidths and improved spectral resolution \cite{PhysRevB.84.195204}. Fig. \ref{fig:6} shows the resulting spectrum in the vicinity of the $\ket{m_s=0}\leftrightarrow\ket{m_s=-1}$ transition. The spectrum exhibits multiple resolved resonance dips, arising from the combined effect of the four crystallographic NV orientations at an applied bias magnetic field of $\approx 1\text{mT}$ and the hyperfine splittings from the intrinsic $^{14}N$ nuclear spin \cite{PhysRevB.79.075203}. The data is fit to a sum of Lorentzian dips (red solid line), from which resonance frequencies, linewidths and relative contrasts of individual transitions were extracted.  
This spectrum directly illustrates the motivation for the amplitude-shaped and adiabatic pulses discussed in Sec.~\ref{sec:theory}: the closely spaced resonances visible here lead to off-resonant excitation with standard microwaves, degrading pulse fidelity. When compounded over the many pulses in a DD sequence, this accumulates into significant coherence loss. Motivated by this, the RAP frequency sweep is centered around one orientation at $2.83876$~GHz, with a R of $7.5$ MHz, spanning the full hyperfine triplet of that orientation. The RAP sweep allows population transfer across this manifold while remaining robust towards the detuning and amplitude inhomogeneity. The extracted linewidths and splittings were used to determine the appropriate microwave drive frequency and to inform the choice of detuning and pulse parameters, ensuring that the applied pulses selectively address the target transition while minimizing unwanted excitation of neighboring resonances.

\begin{figure}[t]
\centering
\begin{minipage}{0.5\textwidth}
  \centering
  \includegraphics[width=\linewidth]{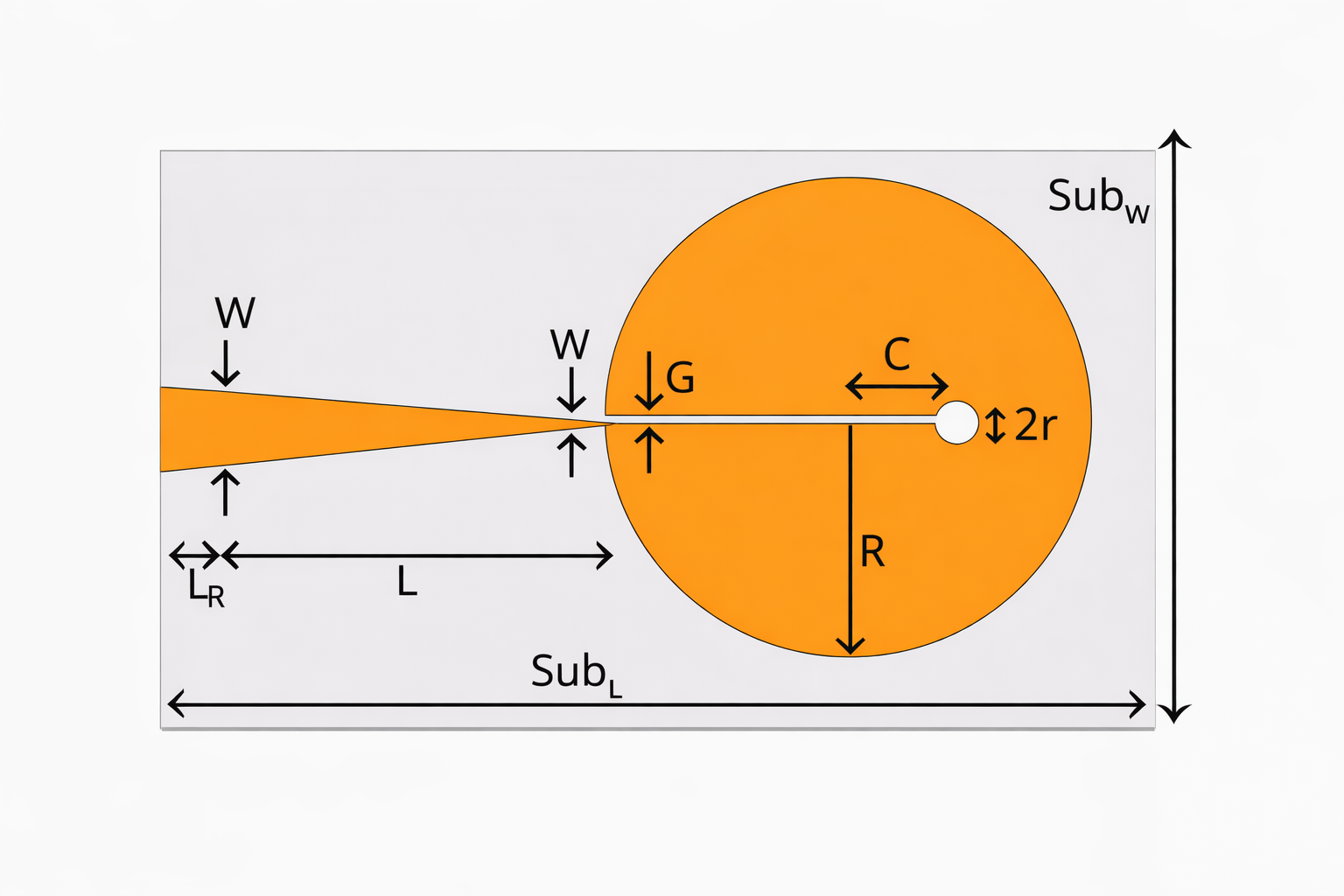}\\
  \small (a)
\end{minipage}\hfill
\begin{minipage}{0.5\textwidth}
  \centering
  \includegraphics[width=\linewidth]{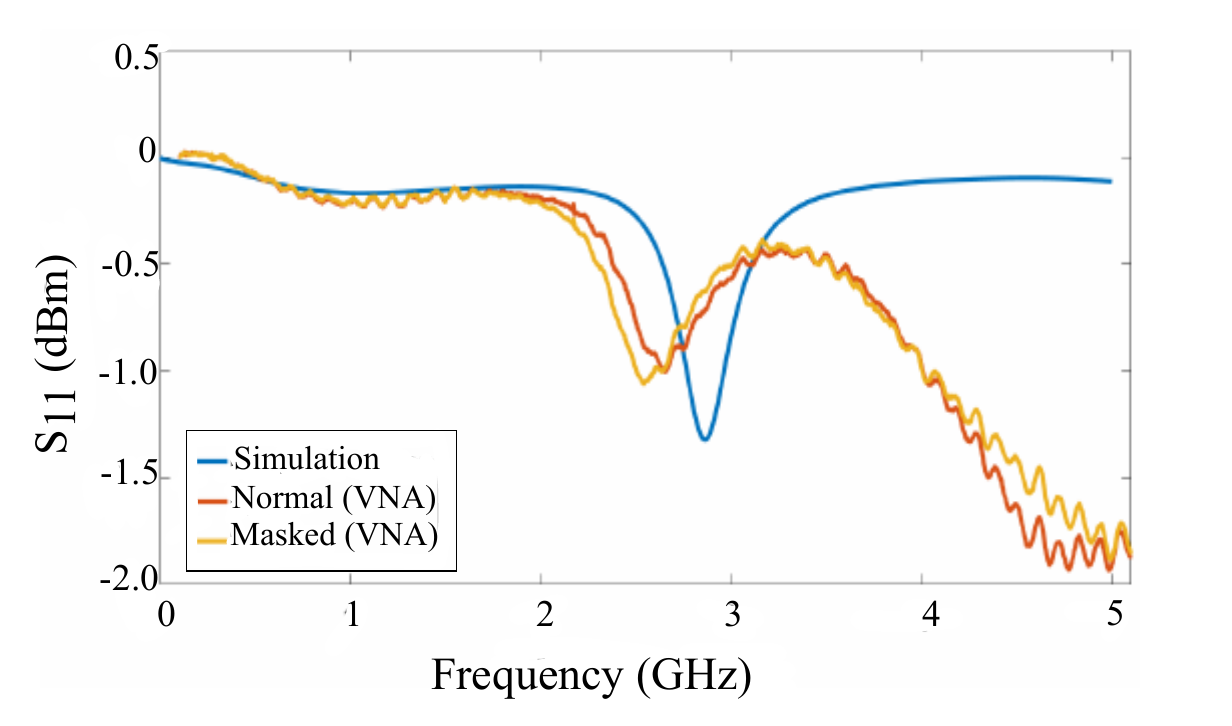}\\
  \small (b)
\end{minipage}
\caption{(a) Schematic of the $LL$ structure. (b) $S_{11}$ parameter for the simulated $LL\_S$ antenna in comparison to VNA measurements (red and yellow curves) of the normal (unmasked) and masked PCBs. The resonance frequency of the $LL\_S$ is $2.651$~GHz with the $S_{11\text{min}}$ value of $-1.005$~dBm measured with the VNA.}
\label{fig:7}\end{figure}

\begin{figure}[t]
\centering
\begin{minipage}{0.5\textwidth}
  \centering
  \includegraphics[width=\linewidth]{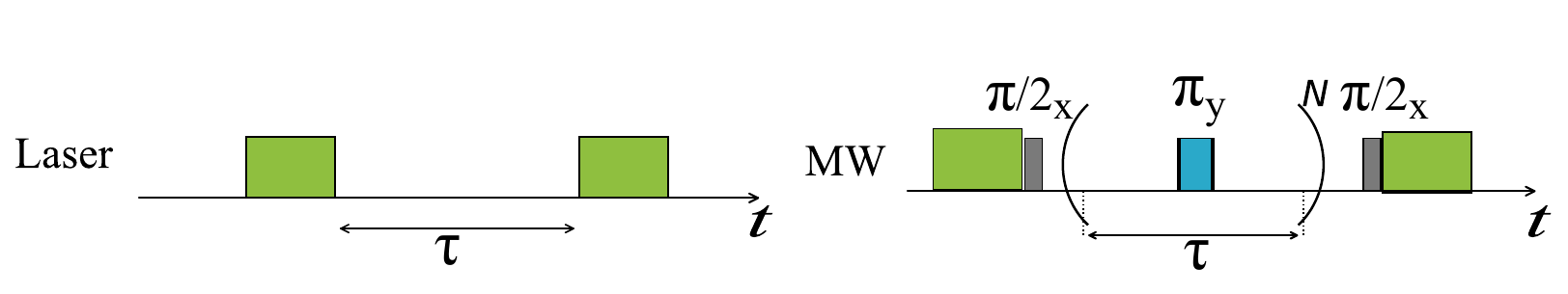}\\
  \small (A)
\end{minipage}\hfill
\begin{minipage}{0.5\textwidth}
  \centering
  \includegraphics[width=\linewidth]{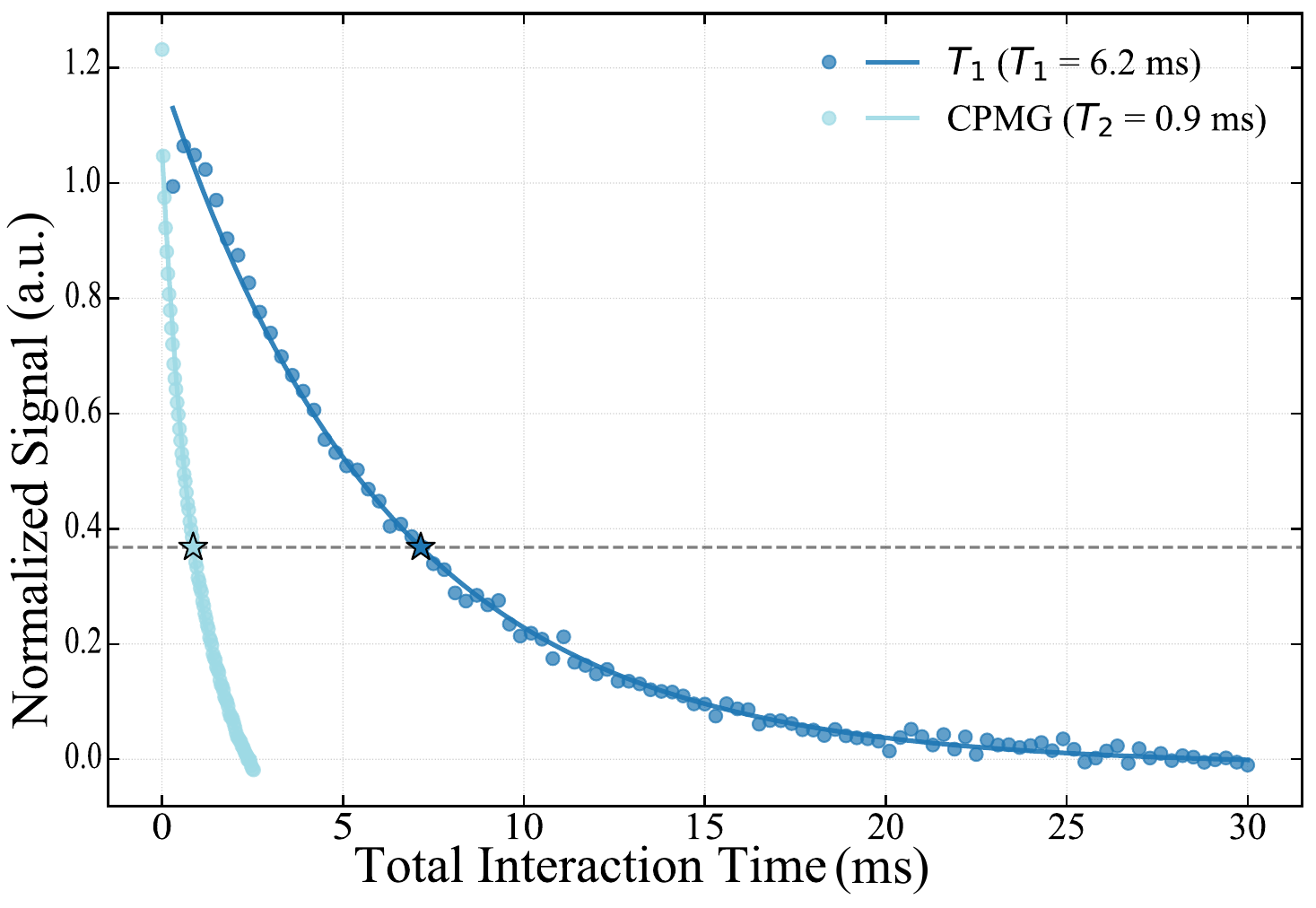}\\
  \small (B)
\end{minipage}
\caption{(A) (Left) Pulse scheme of $T_1$ relaxation time of the ground state $|0\rangle$, where we vary time $\tau$ between the initialization and readout laser pulses in green. (Right) The CPMG sequence, which shows the additional microwave (MW) pulses in comparison to the $T_1$ measurement, where we have a $\pi_y$ pulse in between the two $\pi/2_x$ pulses, with $N$ showing the even number of applied $\pi$ pulses. (B) The CPMG signal is characterized by a decay time to $T_{1,\rho}\approx 900$ $\mu$s for pulse spacing of $\tau=1\,\mu$s. Both the decay curves plotted below fitted with Eq.~\eqref{eqtn:exp}. } 
\label{fig:8}
\end{figure}

\subsection{Microwave antenna Lenz Lens}\label{sec:antenna}
As mentioned earlier, we designed the resonator using simulations and then printed it on the PCB. We modified the dimensions of the Sasaki~\cite{sasaki2016broadband} structure to fit on the cage ($30$ mm Thorlabs Inc.) used for optical mounting of the sample. We refer to these structures as LL (Lenz lens), which include a hole to accommodate the laser beam directed at the sample. Fig.~\ref{fig:7} shows the schematic of the structure.  $L_{R}$ is the length of the untapered part of the microstrip line with the width $W$. $R$ is the radius of the circular main structure, and $r$ is the radius of the hole; the center of the hole is displaced by $C$. The gap in the copper coating, from the end of the micro strip line to the hole, has a width of $G$. We developed three structures based on the value of $R$, named accordingly: the largest structure is called $LL\_L$ ($R=10$~mm); a medium size $LL\_M$ ($R=8$~mm); and the smallest $LL\_S$ ($R=7$~mm). For each, we developed both a masked and an unmasked variant. We performed the experiment with $LL\_S$ owing to its least inhomogeneity in the Rabi drive among all the structures.

\subsection{RF resonator}
We also developed the RF resonator based on the omega-loop structure based on~\cite{bayat2014efficient} to deliver the test signal. Because the radius of this structure is larger than that of the microwave resonator, we placed the PCB below the RF resonator to accommodate both. We characterized this combined structure with a vector network analyzer (VNA) to check for mutual interference. The $S_{11}$ parameters of both resonator ports showed no significant degradation, indicating that the Rabi drive was unaffected by the presence of the RF resonator; we further confirmed this directly by performing Rabi oscillation measurement on the combined structure.

\begin{figure*}[t]
\centering
    \includegraphics[width=\textwidth]{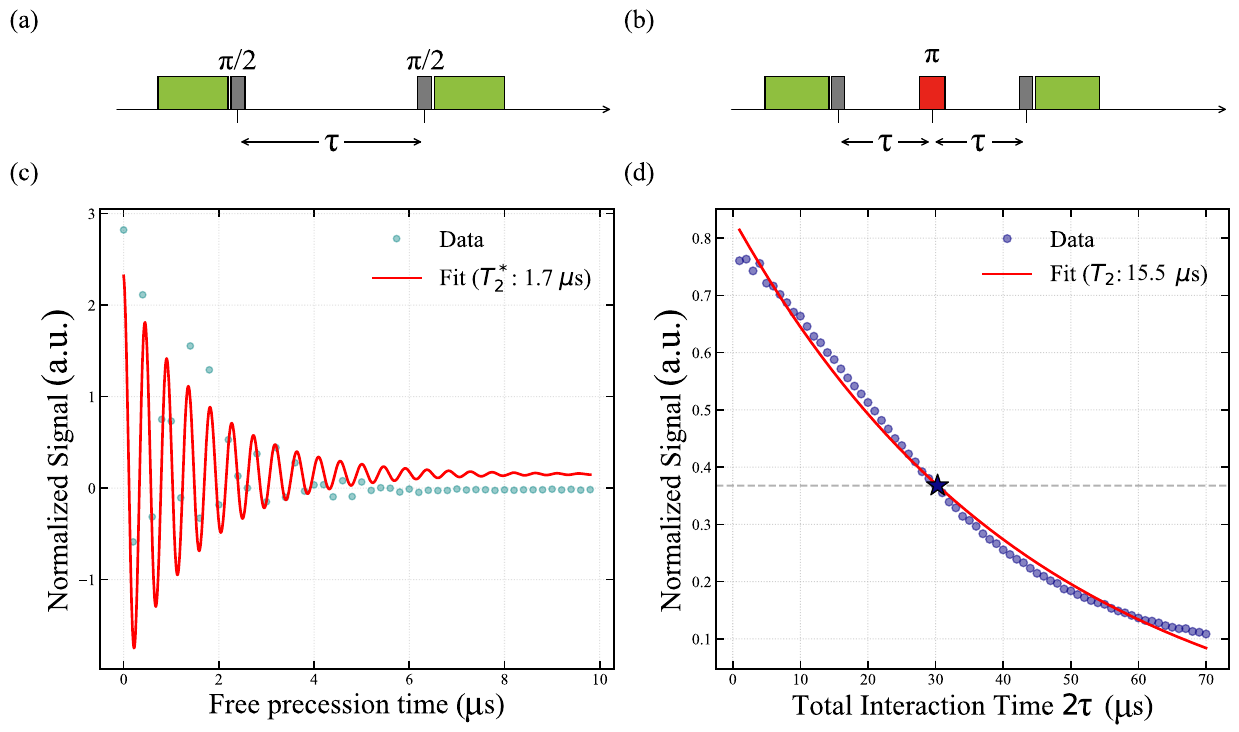}
\caption{(a) Pulse schemes for Free Induction decay or Ramsey protocol and (b) Hahn echo protocol. (c) The Ramsey signal decay curve fitted with a damped cosine function: $a e^{-t / T_2^*} \sin(2 \pi f t + \phi) + c$ where a is the amplitude, f is the frequency of oscillation, $\phi$ is the phase, and c is an offset. $T_2^*$ is the dephasing time.(d) The decay curve fitted with an exponential function, following Eq.\eqref{eqtn:exp} from the Hahn echo measurement shows that the signal decays to $1/e$ of the normalized Rabi signal at $t=2\tau=31 \pm 1.1\,\mu$s. Thus, the Hahn echo decay time, which corresponds to the $\tau$ when the signal decays to this value is $T_2=15.5 \pm 0.6\,\mu$s.}
\label{fig:9}
\end{figure*}

\begin{figure}[t]
\centering
\includegraphics[width=\columnwidth]{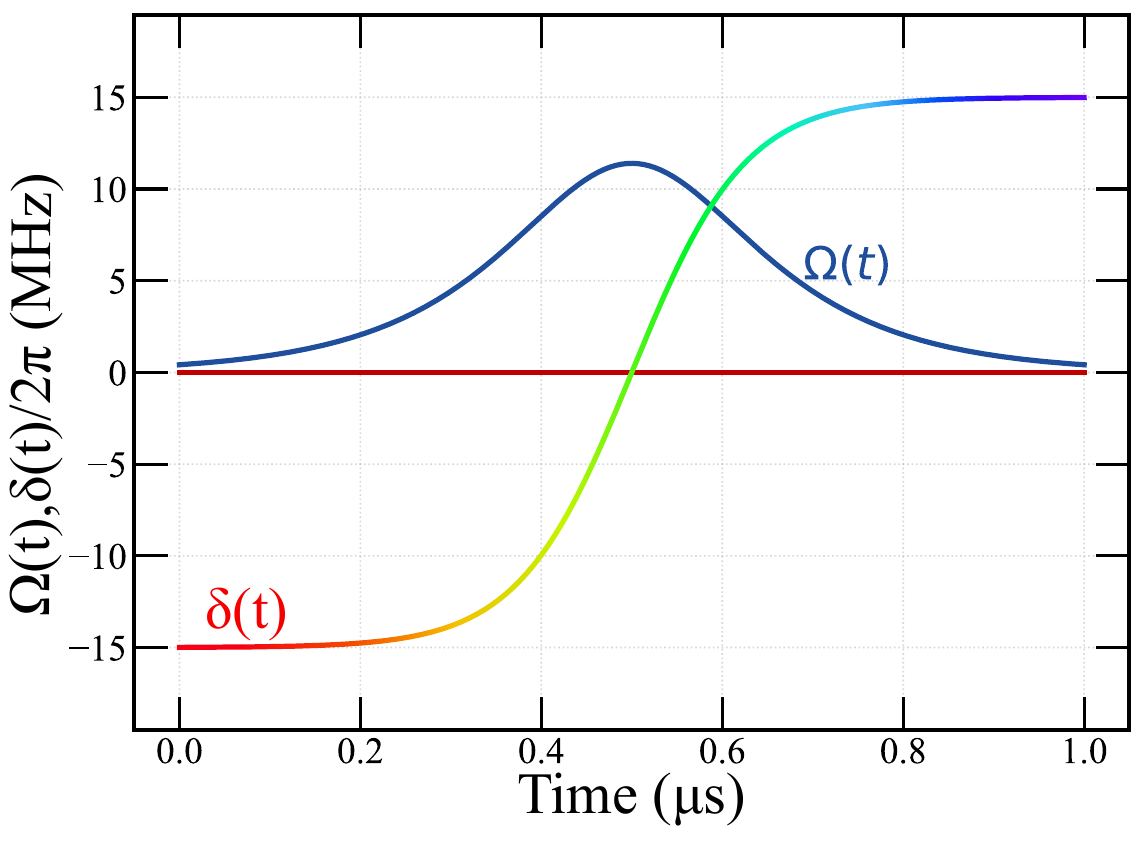}
\caption{RAP pulse shape used in the experiments. The Rabi-frequency envelope $\Omega(t)/2\pi$ (blue) follows a hyperbolic-secant profile, while the detuning sweep $\delta(t)/2\pi$ follows a hyperbolic tangent, crossing zero at the pulse center $t_c=0.5\,\mu\text{s}$. Both are shown for the experimentally optimized parameters, $T_{\text{pulse}}=1\,\mu\text{s}$
and one of the detuning range of $R=(2\pi)30$~MHz.}
\label{fig:rap_shape}
\end{figure}

\section{Additional measurements}\label{Sec:App_B}

\begin{figure*}[htbp]
\centering

% --- Top row: (a) and (b) ---
\begin{minipage}{0.48\textwidth}
  \centering
  \small (a) \\
  \includegraphics[width=\linewidth, keepaspectratio]{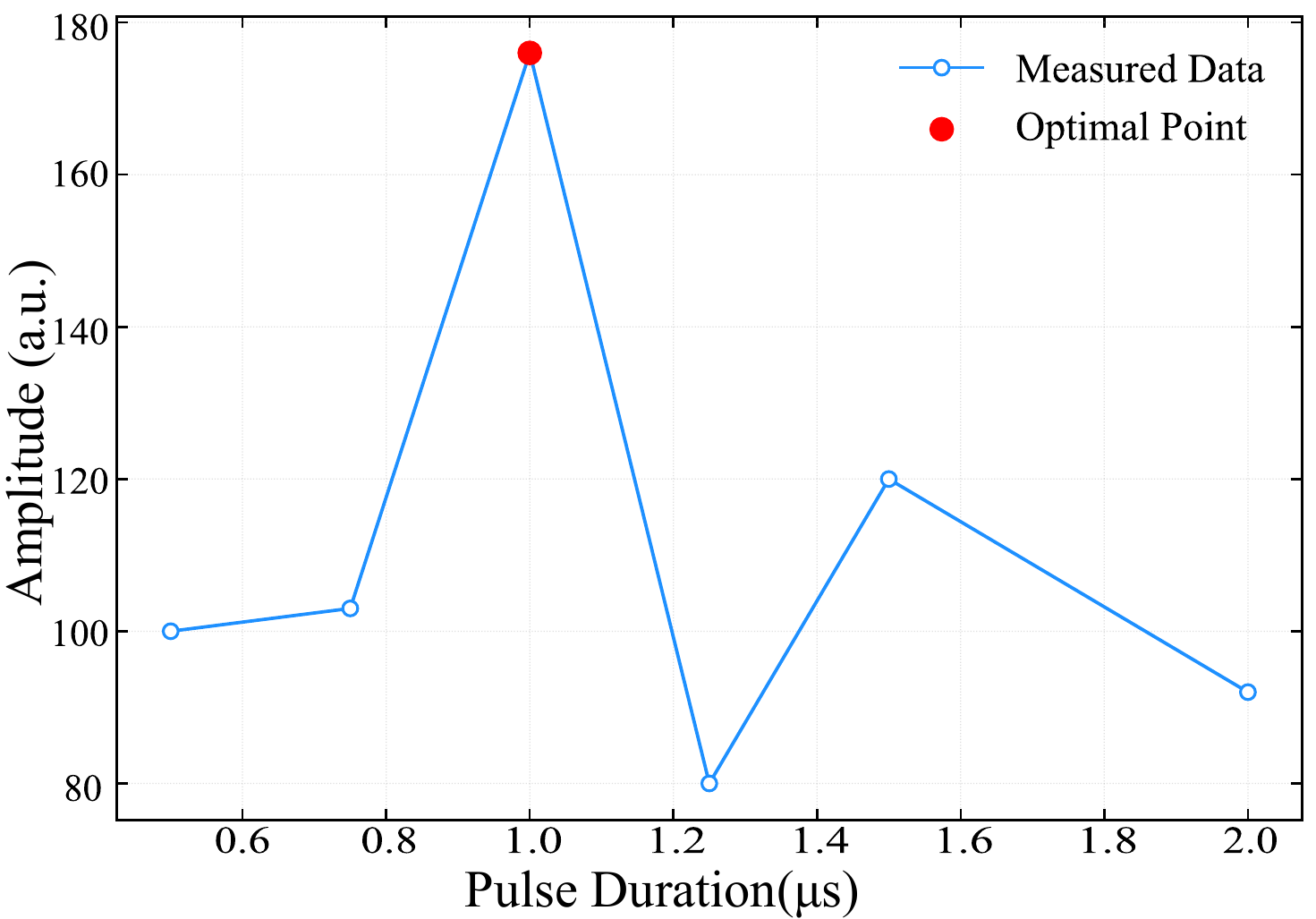}
\end{minipage}
\hfill
\begin{minipage}{0.48\textwidth}
  \centering
  \small (b) \\
  \includegraphics[width=\linewidth, keepaspectratio]{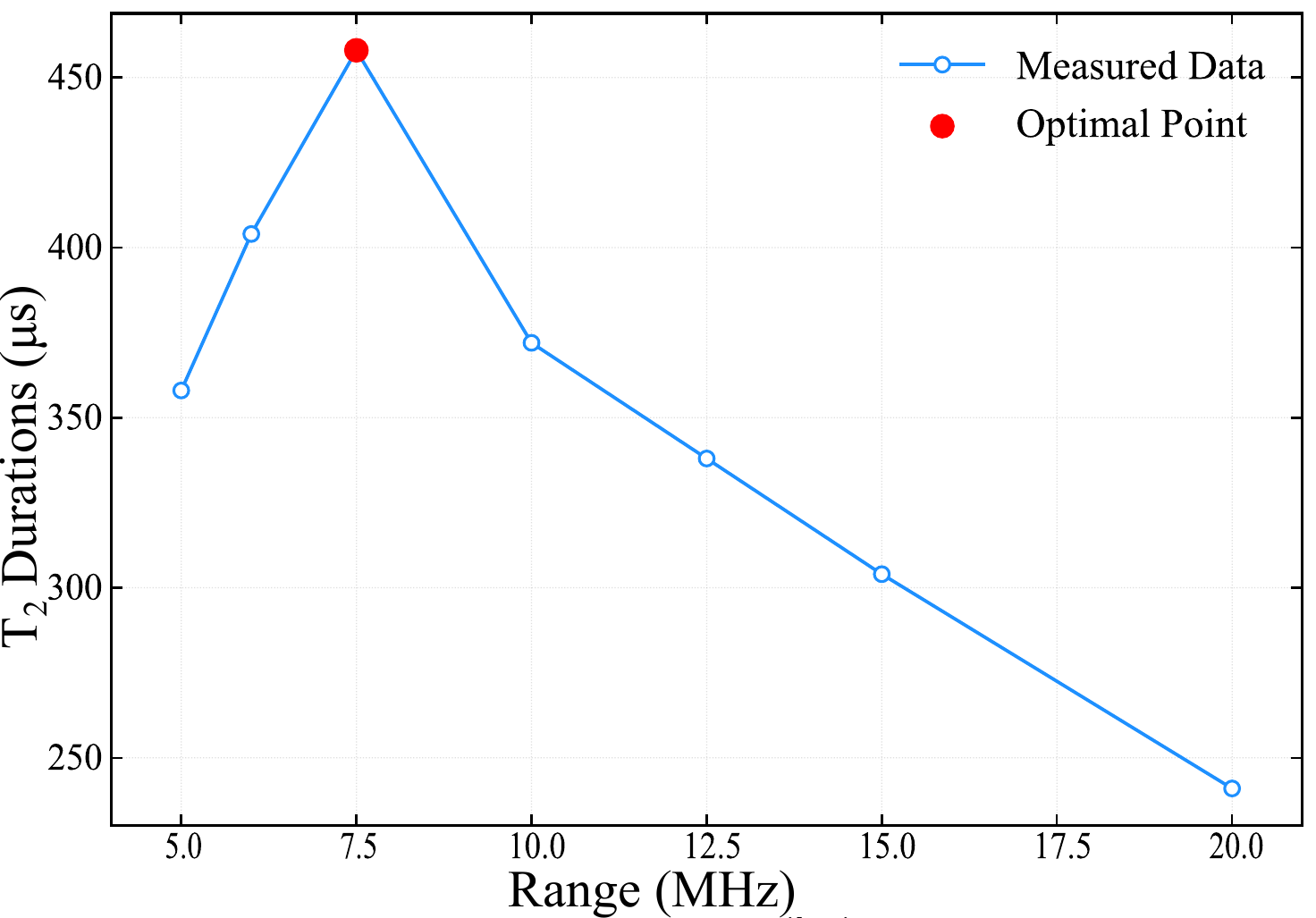}
\end{minipage}

\vspace{0.3cm}

% --- Bottom row: (c) and (d) ---
\begin{minipage}{0.48\textwidth}
  \centering
  \small (c) \\
  \includegraphics[width=\linewidth, keepaspectratio]{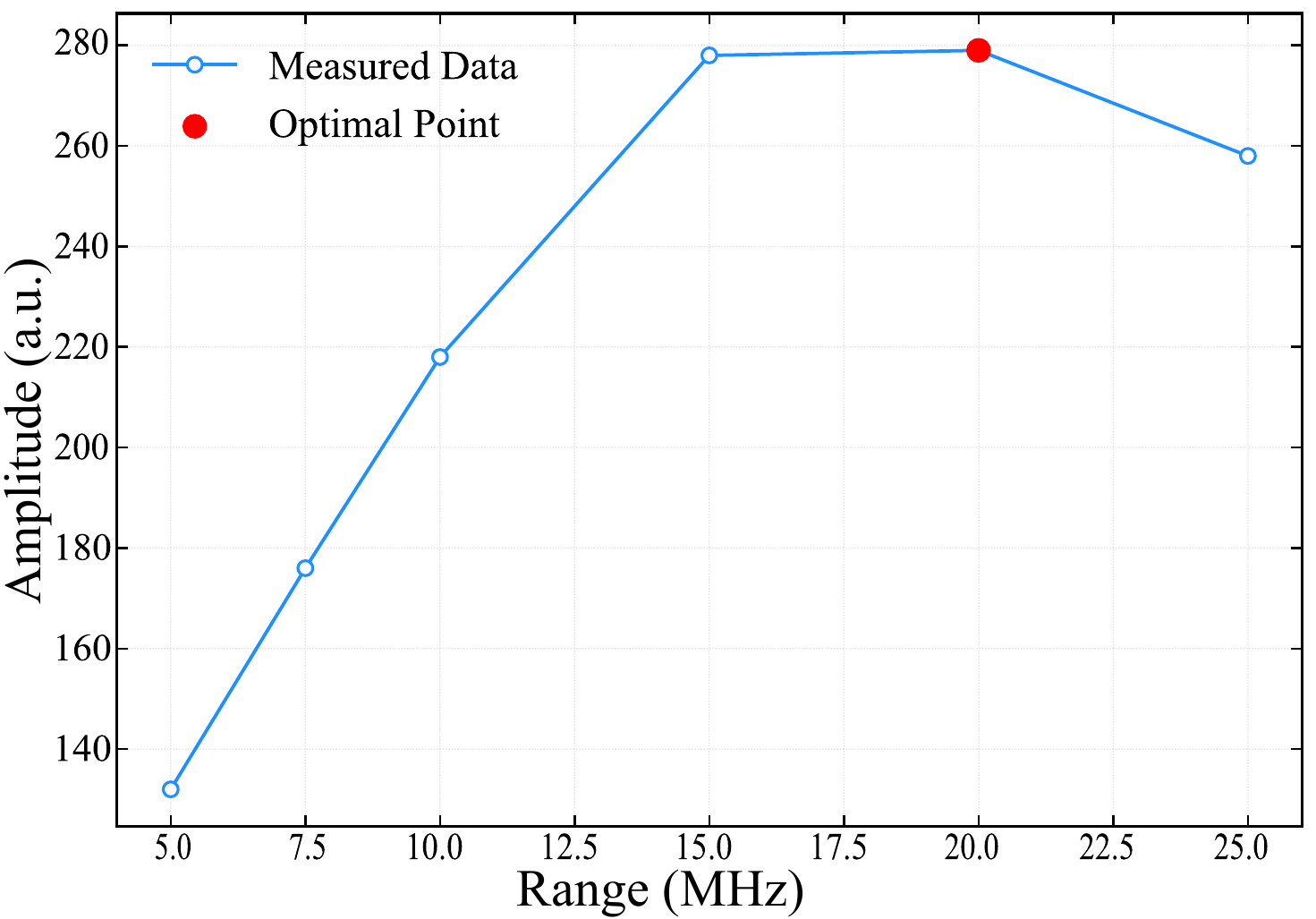}
\end{minipage}
\hfill
\begin{minipage}{0.48\textwidth}
  \centering
  \small (d) \\
  \includegraphics[width=\linewidth, keepaspectratio]{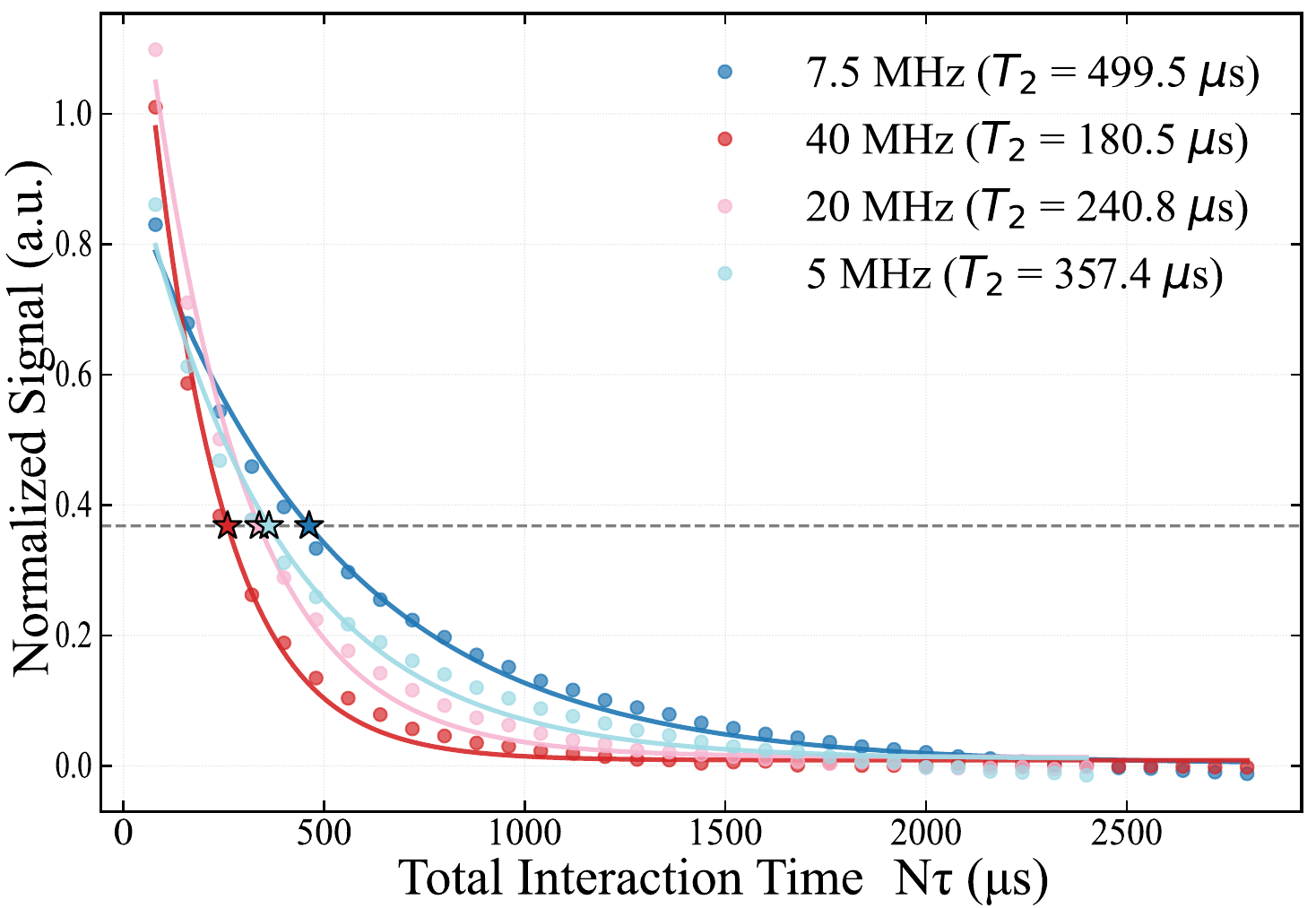}
\end{minipage}

\caption{Optimization of RAP parameters. (a) Pulse duration optimized at $1 \mu s$ for KDD16 with range ($R=7.5$~MHz), this optimal duration holds consistently across range values from $R=5$ to $25$~MHz. (b) Range optimization with respect to $T_2$, keeping $T_{\text{pulse}}=1 \mu s$ fixed; the optimal value is found to be $R=7.5$~MHz. (c) Range optimization with respect to initial signal amplitude, showing that higher range values, such as $R=20$~MHz, give a larger initial contrast. (d) Decay curves, where we vary $N$, while $\tau=T_{\text{pulse}}=1 \mu s$ showing that longest $T_2$ is obtained at $7.5$ MHz, while signals at higher range values exhibit better initial contrast but decay faster, as indicated by the stars at $1/e$ (on the dashed horizontal line). All decay curves are fitted to the exponential function given in Eq.~\eqref{eqtn:exp}}.
\label{fig:10}
\end{figure*}

\subsection{Measurements for $T_1$, CPMG $T_{1\rho}$ Limit}
We measure the $T_1$ relaxation time by initializing the system with a laser pulse into state $|0\rangle$, as shown in Fig. \ref{fig:1} in the main text, and waiting for time $\tau$. Another laser pulse is then used as a readout. We mark the relaxation time to the thermal equilibrium of the system with the decay of the curve. 
The limit for dynamical decoupling lifetimes is determined by the CPMG scans of higher order ($N$) Fig.~\ref{fig:8}. The experimental data were fitted using a single exponential decay model:
\begin{equation}
    S(t) = a e^{-t/T_2} + c 
    \label{eqtn:exp}
\end{equation}
where $S(t)$ represents the normalized signal to the Rabi measurement at interaction time $t$, $a$ is the initial amplitude and $c$ is a constant offset accounting for the steady-state background. The decay constant $T_2$ represents the coherence time, which corresponds to $T_{1,\rho}=N\tau$ for CPMG for the particular pulse separation of $\tau=1\,\mu$s. The star marks on the decay plots signify the $1/e$ levels. 

\begin{figure*}[t]
\centering
\includegraphics[width=\textwidth]{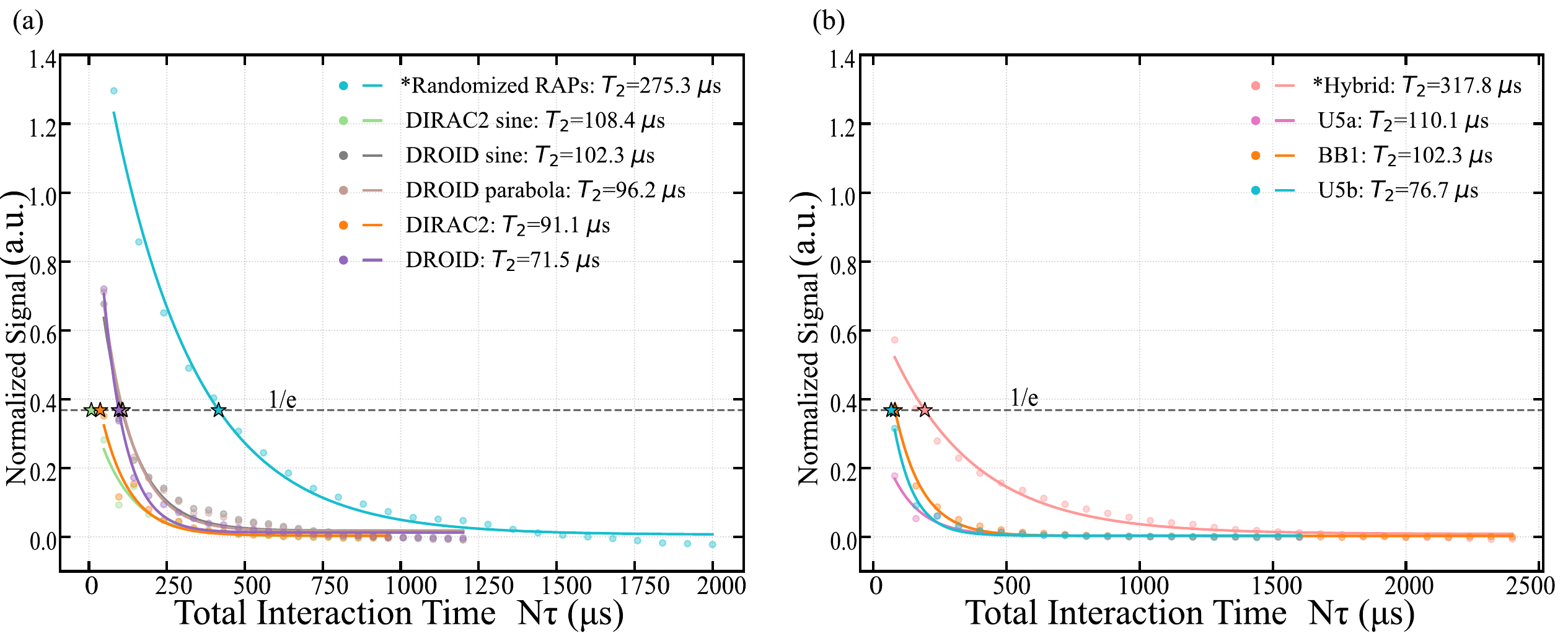}
\caption{(a) The decay signals of the DROID scans are plotted against the total time evolution, where $N$ is the total number of $\pi$ pulses and $\tau$ is the inter-pulse delay in the sequence. (b) Dynamical decoupling with the standard Composite Pulses compared to Hybrid RAPs.} 
\label{fig:11}
\end{figure*}

\subsection{Free Induction Decay and Hahn Echo Measurements}
To measure the dephasing time ($T_2^*$) of the sample, we perform the Ramsey protocol. The initialized spin state $m_s = 0$ is brought through a $\pi/2$ pulse in a superposition state. The time $\tau$ in that state is varied, and finally, another $ \pi/2$ pulse converts the coherence into population, allowing the state to be read out with a laser. The same is done with a $\pi$ pulse in between the two $\pi/2$ pulses to determine the decoherence time $T_2$. The measurement data along with the pulse scheme is depicted in~\ref{fig:9}.

\subsection{Pulse Envelope Shaping for Dynamical Decoupling}
\label{Subsec:Pulse_shaping}
In addition to the standard ($\pi$- and $\pi/2$-) pulses, we implemented DD sequences using parabolic (with order $p=1,2,3$) and $\sin^n$ (with $n=1,2,3$) shaped pulse envelopes, motivated by the reduction of power broadening achievable with smooth, finite-duration pulse-envelopes~\cite{BoradjievPRA2013,MihovPRA2024}. 
For a two-level system, driven by pulses of finite duration $T$ and shape function say $E(t)$, with a Rabi frequency of $\Omega(t) = \Omega_0 E(t)$, the excitation linewidth (the width of the transition probability $P_{0\to1}$ as a function of detuning $\delta$) is governed primarily by the order $n$ of the first discontinuous derivative of $E(t)$ at the beginning and end of the pulse duration $t=0$ and $t=T_{\text{pulse}}$ \cite{BoradjievPRA2013}. For a standard $\pi$ pulse, $\Omega_0 T_{\text{pulse}} = \pi$ and $E(t)$ itself is discontinuous($n=0$), resulting in linear power broadening of the Rabi linewidth $\delta_{1/2}\propto \Omega_0$. For pulse shapes with a smooth, $n$-times differentiable start/end, the non-analyticity manifests as a singular coupling in  the $n{+}1$-th superadiabatic basis, yielding substantially weaker power broadening \cite{BoradjievPRA2013},
\begin{equation}
    \delta_{1/2}\propto \Omega_0^{1/(n+1)}.
\label{eqtn:power_broad}
\end{equation}
\paragraph{Sine- and parabola-power envelope.---} To design the shaped envelopes, we define the normalized time coordinate $x=t/T_{\text{pulse}} \in [0,1]$\cite{BoradjievPRA2013} so that
\begin{equation}
    E_n(x)=\sin^n(\pi x), \qquad n \in \mathbb{N},
\label{eqtn:sine}    
\end{equation}
with $\Omega(t)= \Omega_0 E_n(t/T_{\text{pulse}})$. And using the symmetric variable  $u(x)=2x-1 \in [-1,1]$, the parabolic-power envelope is
\begin{equation}
    E_p(x)=1-u(x)^{2p}=1-(2x-1)^{2p},\qquad p \in \mathbb{N},
\label{eqtn:parabola}    
\end{equation}
Both envelopes vanish smoothly at the edges, $E(0)=E(1)=0:sin^n{\pi x}$ has a discontinuous $(n+1)$-th derivative, and the $p=1$ parabola has a discontinuous second derivative, placing it in the smoothness class analogous to $n=2$.
For every envelope, the peak Rabi amplitude $\Omega_0$ was fixed and the pulse duration $T_{\text{pulse}}$ (equivalently, the Rabi period) was set by requiring a constant pulse area, 
\begin{equation}
\int_0^{t} \Omega(t^\prime),dt^\prime = \pi,
\label{eqtn:area_condition}
\end{equation}
evaluated at $t=T_{\text{pulse}}$, so that each pulse implements the same $\pi$-pulse (population inversion) at the same drive power. Because a shaped envelope integrates to less than the flat-top value over $[0,T_{\text{pulse}}]$, this fixed area condition requires a somewhat longer Rabi period for the shaped pulses relative to the standard ones. This is consistent with our measurements: the calibrated Rabi periods measured were $T_{\text{pulse,std}}=91.74$ ns, $T_{\text{parabola,p=1}}=112.23$ ns and $T_{\text{sine,n=1}}=116$ ns.

For both envelope families, the full shaped microwave drive applied to the qubit is
\begin{equation}
\Omega_{\text{shaped}}(t) = 2A_0 \sin\!\big(\omega_0 t + \phi\big)\, E(t/T_{\text{pulse}}),
\label{eqtn:shaped_pulse}
\end{equation}
with carrier amplitude $A_0$, carrier frequency $\omega_0$, and phase $\phi$ as defined in Eq.~\eqref{eq:1}, and $E(t/T_{\text{pulse}})$ given by Eq.~\eqref{eqtn:sine} or Eq.~\eqref{eqtn:parabola}.\\
This reduction in power broadening is particularly relevant to our experiments, which were performed at a low bias magnetic field, where the Zeeman splitting between the $\ket{m_s=\pm1 }$ ground states is of the order of tens of MHz. There are also multiple, closely spaced spin transitions as the nuclear spin of the nitrogen is not polarized and there could be NV centers with different orientations. Consequently, a standard pulse driving $\ket{m_s=0} \leftrightarrow \ket{m_s=+1 \text{ or } -1}$ can leak the population into these nearby off-resonant levels due to its broad excitation profile, degrading fidelity and in a DD sequence, accumulating coherent errors over several pulses.  The reduced power broadening of the sine- and parabola-power envelopes narrows the excitation profile around the target level, suppressing this leakage and thereby mitigating unwanted off-resonant driving of neighboring spin states.
\section{RAP optimization}\label{sec:opt_raps}
%\subsection{Optimizing the RAPs}
Here we present additional details on the experimental optimization of the RAP parameters. 
We apply amplitude and frequency modulation of the RAP pulse, as shown in Fig. \ref{fig:rap_shape}, which follow the Allen-Eberly model \cite{Allen-Eberly1987,genov2020efficient}
\begin{equation}
    \Omega(t) = \Omega_0\,\text{sech}\!\left(
    \frac{t-t_c}{T}\right),~
    \delta(t) = (R/2)\,\tanh\!\left(
    \frac{t-t_c}{T}\right),
\label{eqtn:rap_pulse_shape}
\end{equation}
where the pulse is centered at $t_c$, $\Omega_{0}$ is the peak Rabi frequency, and $R$ is the chirp range. 
The amplitude modulation was also controlled via the truncation ratio, $T_{\text{pulse}}/T$, which was optimized and identified to be $T_{\text{pulse}}/T\approx8$. Here, $T$ is the characteristic time of the underlying hyperbolic secant Rabi-frequency envelope and hyperbolic tangent detuning sweep. 
The characteristic time $T$ sets the timescale over which these functions may vary: at $t-t_c=\pm T$ the hyperbolic secant (sech) envelope has fallen to $\text{sech}(1)\approx0.648$ of its peak value, so shorter $T$ corresponds to a more sharply peaked pulse, while larger $T$ value corresponds to a more gradual variation. Since ideal sech/tanh functions extend to $t=\pm\infty$, while the applied pulse is truncated to the finite window $t\in[t_c-T_{\text{pulse}}/2,\,t_c+T_{\text{pulse}}/2]$, the truncation ratio $T_{\text{pulse}}/T$ quantifies how many characteristic times fit within this window. A larger $T_{\text{pulse}}/T$ means the envelope has already decayed substantially by the window edges, resulting in a smaller residual discontinuity upon truncation at the expense of a longer pulse. In contrast, a smaller $T_{\text{pulse}}/T$ makes the pulse duration shorter but truncates the pulse closer to its peak, introducing a more significant discontinuity and, correspondingly, a larger deviation from ideal adiabatic behavior.  

As a next step, we optimized the frequency chirp range ($R$) and the pulse duration $T_{\text{pulse}}$ of the RAPs by comparing the resulting signal contrast and $T_{2}$ times across several values. 
Figure~\ref{fig:10} shows that $T_{\text{pulse}}=1\mu s$ and $R=7.5$~MHz yield the optimal $T_{2}$ time. Although the initial contrast is higher at larger chirp ranges, such as $20$~MHz, the corresponding signals decay considerably faster than those obtained with $R=7.5$~MHz. This trade-off is illustrated by the star markers at the $1/e$ decay point in each curve in Fig. \ref{fig:10}(d), indicating the coherence decay to $1/e$ of the contrast of the signal, normalized to a Rabi oscillations measurement, for each chirp range.

\section{Additional details on comparison with other dynamical decoupling sequences}\label{sec:other_DD_methods}

In order to compare the performance of advanced dynamical decoupling methods that compensate both disorder, and unwanted spin-environment and spin-spin interactions, we also employed higher-order methods for Hamiltonian engineering pulse sequences including DROID and DIRAC2~\cite{tyler2023higher}. In addition, we also used randomized pulse-phase sequences \cite{wang2019randomization} for comparison, and implemented shaping of the pulses while applying these methods. In Fig.~\ref{fig:11} (a), we compare the memory time obtained with DROID, DIRAC2, DIRAC2-sine, DROID-sine, and DROID-parabola sequences to that of randomized-phase KDD16-RAP scans, where each randomized-phase run uses an independently drawn random phase set, and the reported curve is the average over 30 runs. The randomized-phase RAPs outperform the DIRAC2 and DROID variants, though they still do not outperform the optimized (non-randomized-phase runs) RAP sequences. In Fig.~\ref{fig:11}(b), we show the performance of composite pulses  U5a, BB1, and U5b (see main text for references), compared with hybrid RAP sequences, in which $\pi$ pulses were implemented as RAPs while the initial and final $\pi /2$ pulses remain standard. The hybrid RAP sequences yield the best performance among all composite-pulse variants tested, apart from the RAP sequences in the main text, which also applied half-RAP pulses instead of standard, rectangular $\pi/2$ pulses. Decay times marked with stars indicate the $1/e$ point of each curve. All decay curves were fit to the  single exponential fit function of the form in Eq.\eqref{eqtn:exp}.

\begin{figure*}[!t]
     \centering
     \begin{subfigure}[!t]{0.48\textwidth}
         \centering
         \includegraphics[width=\textwidth]{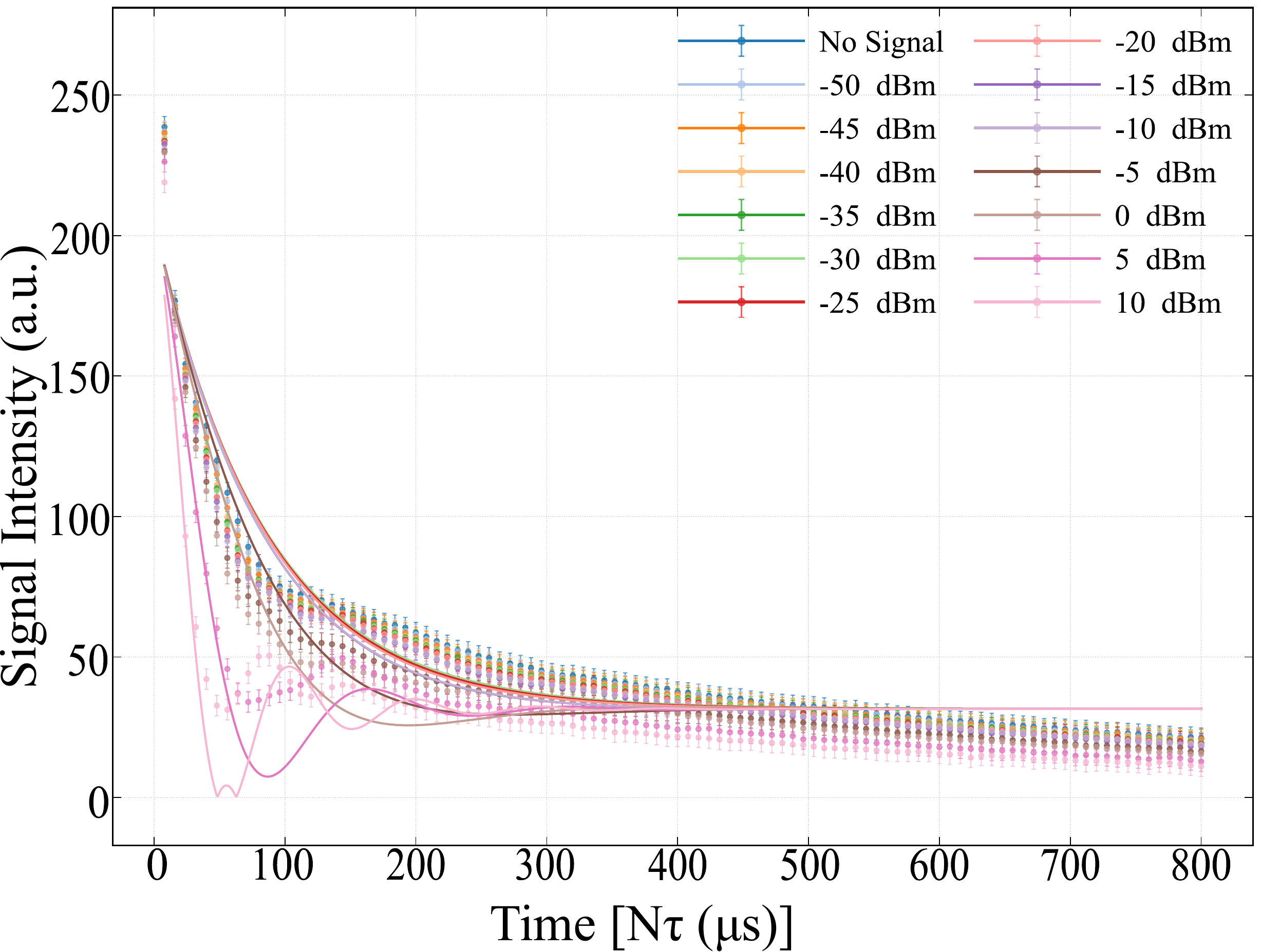}
         \caption{XY8 order sweep with external ac signal.}
         \label{fig:sub1}
     \end{subfigure}
     \hfill
     \begin{subfigure}[!t]{0.48\textwidth}
         \centering
         \includegraphics[width=\textwidth]{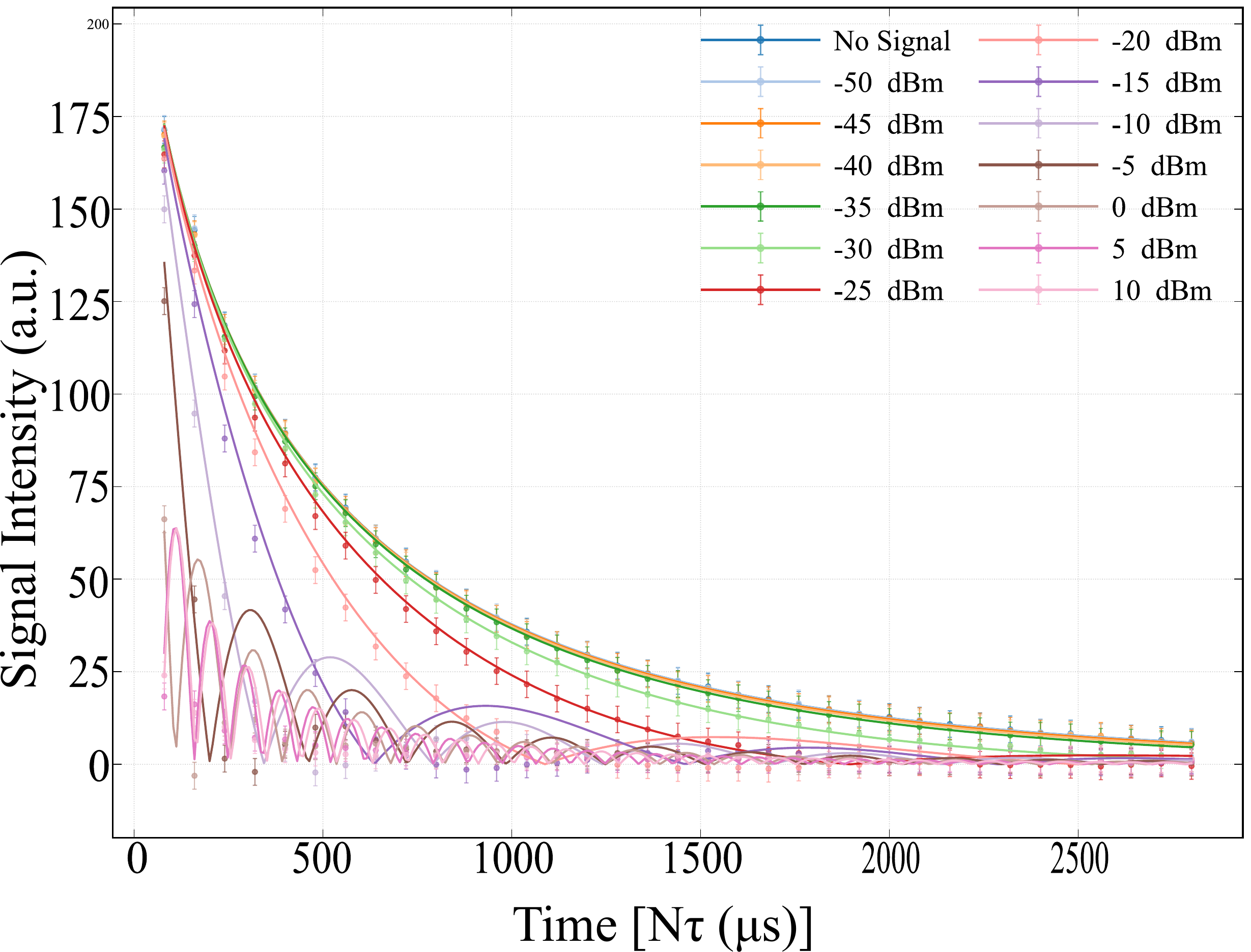}
         \caption{KDD16 order sweep with the same probe Rf signal.}
         \label{fig:sub2}
     \end{subfigure}
        \caption{Comparison between XY8 ($\tau=1 \mu\text{s}$ since $f_p = 1/2\tau$) and KDD16 sequences with fixed pulse duration $T_{\text{pulse}}= 1 \mu s$, while a $f_p=500$ kHz RF signal is applied. The resulting NV fluorescence response is fit using equation \eqref{eqtn2}.}
        \label{fig:13}
\end{figure*}

\section{Additional experimental results: quantum sensing}
\subsection{Amplitude Sensing/AC Magnetometry}\label{sec:amplitude_details}
Building on the improved coherence properties demonstrated above, we extended the RAP based DD protocol to perform AC magnetometry. Our first goal is to sense the amplitude `'g'' of an oscillating classical non-phase-matched signal with an angular frequency $\omega_s=(2\pi)f_s$, similarly to \cite{genov2020efficient}. 
To test the sensitivity of the setup, we introduce an external ac field with the goal of measuring its amplitude. 
We vary the amplitude of the signal and apply the KDD16-RAP sequence, where we vary the order $N$ of the sequence, i.e., the number of times it is repeated, detecting the relevant signal changes due to the presence of the target sensed field. As we vary $N$, we keep the time between the centers of the RAP pulses constant, i.e., $\tau=T_{\text{pulse}}+\tau_{\text{ip}}$. The pulse duration is selected so that the sequence's filter function is resonant with the external signal frequency, via $\tau=T_{\text{pulse}}=1/(2f_s)=\pi/\omega_s$. Here the external frequency was set to $f_s=500$~kHz with $T_{\text{pulse}}=1 \mu\text{s}$. We fit the signal with:
\begin{equation}
    S(t) = c + a J_0(2\pi g  t) e^{-\left({t}/{T_2}\right)^{\beta}}
    \label{eqtn2}
\end{equation}
where the \enquote{g} parameter is the coupling strength through which the signal from the order-$N$ scan is coupled to the external signal. 
The fitting was performed in two stages. First, a reference \enquote{no signal} dataset was fitted to a standard stretched exponential decay ($g=0$) to determine the intrinsic decoherence parameters, including  $T_2$ (decoherence time), $\beta$ (the stretching exponent describing the noise environment), and the offset \enquote{c}.
\enquote{a} is the initial signal contrast and \enquote{$J_0$} is the zeroth-order Bessel function of the first kind, accounting for the coherent phase accumulation of the NV center due to the external RF field.  These parameters were then used to constrain the fit for the signal data, where the coupling parameter \enquote{g} was extracted using the NonlinearModelFit function of Mathematica with the option  (\texttt{NMinimize}) to find the parameters that minimize the sum of squared residuals of the Bessel function model (Eq. \ref{eqtn2}). Error bars in the figures (Fig. \ref{fig:13}) represent the standard deviation of the fluorescence signal, and the quality of the fit was assessed using the adjusted $R^2$ values.
To validate the physical consistency of the estimated coupling parameters, we compared the extracted values of $g$ against the expected amplitude of the applied RF field. The applied RF power, measured in dBm, is converted to an amplitude factor, $A_{\text{RF}}$, which is proportional to the square root of the power in milliwatts:
\begin{equation}
    A_{\text{RF}} \propto \sqrt{10^{(P_{\text{dBm}}/10)}}
    \label{eq:dbm_to_amp}
\end{equation}
where $P_{\text{dBm}}$ is the set power of the RF source.
Since the coupling strength $g$ is expected to be linearly proportional to the field amplitude ($g\propto B_{\text{ac}}\propto A_{\text{RF}}\propto\sqrt{P}$, via Eq.~\eqref{eq:dbm_to_amp})~\cite{taylor2008,degen2017}, we performed a normalization procedure to compare results across different power levels. To test this scaling, we define the normalized coupling ratio,
\begin{equation}
    \mathcal{R}_{g,i} = \frac{g_i}{g_{\text{ref}}}
    \label{eq:abs_ratio}
\end{equation}
where $g_i$ is the estimate for a given power level and $g_{\text{ref}}=g_{0\text{dBm}}$ is the value obtained at the reference power of $0$~dBm. The corresponding expected ratio is computed as $A_{\text{RF},i}/A_{\text{RF},\text{ref}}$ using the same normalization, so that the unknown proportionality constant in Eq.~\eqref{eq:dbm_to_amp} cancels out.
By comparing the experimental $\mathcal{R}_{g,i}$ values against the theoretically expected amplitude ratios derived from Eq.~\eqref{eq:dbm_to_amp}, we can identify any potential saturation effects or deviations from linearity in the NV
ensemble's response. As shown in Fig.~\ref{fig:3}(lower) in the main text, the estimated values exhibit a strong linear correlation with the expected scaling ($y=x$). This agreement confirms that the Bessel-model fit accurately captures the
phase accumulation dynamics of the spin ensemble across several orders of magnitude of signal power, from $-50$~dBm to around $0$~dBm, with higher power values showing effects of saturation. The signals of the sensing protocols using standard XY8 and KDD16-RAP (with $R=7.5$~MHz) fitted with
Eq.~\eqref{eqtn2} are illustrated in Fig.~\ref{fig:13}.

\begin{figure}[!t] 
    \centering
    \includegraphics[width=\columnwidth]{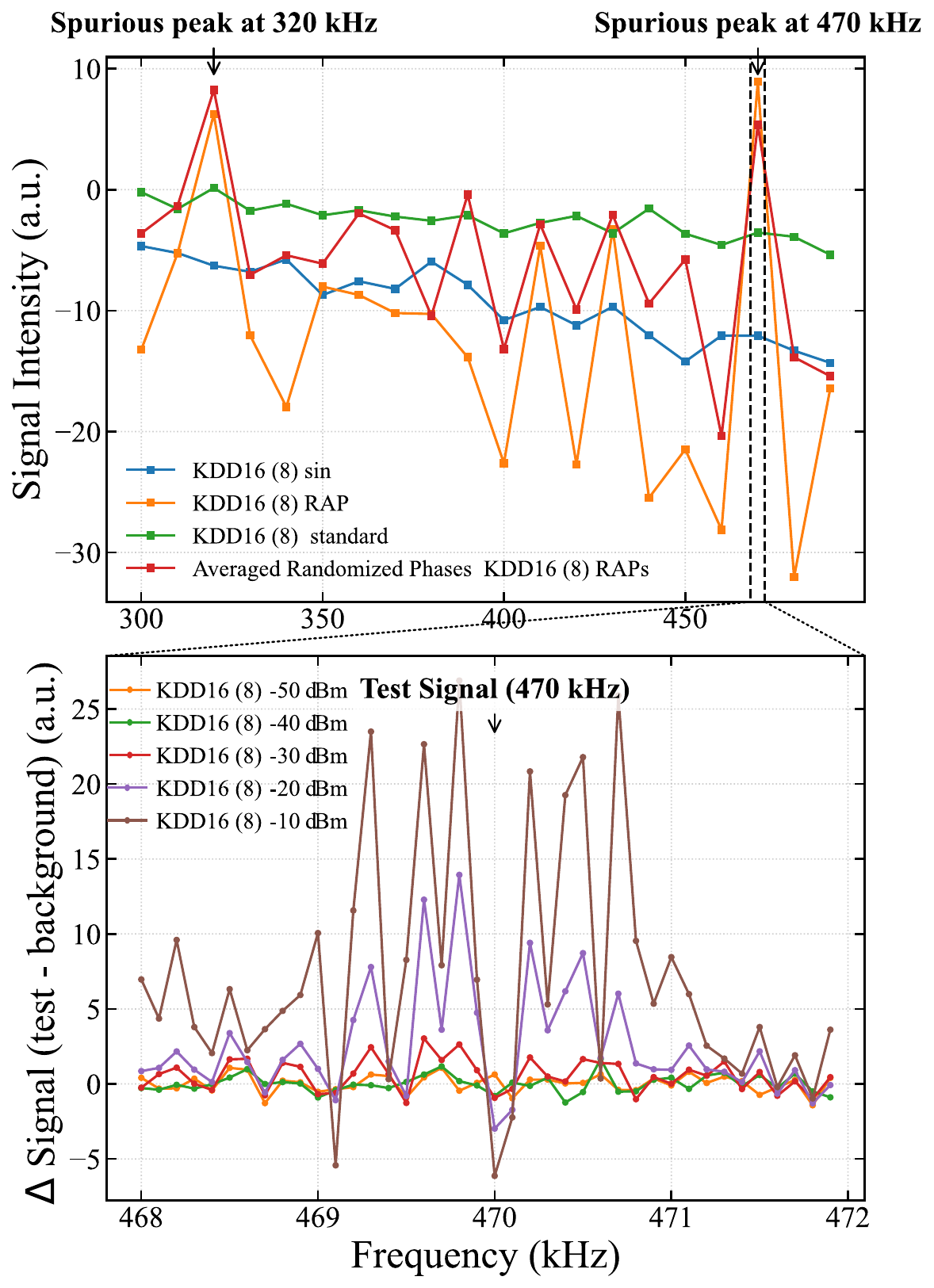}
    \caption{Representation of the frequency-domain characterization of the KDD16 sensing scheme.  The upper plot shows the background signal spectra (no external test field applied) obtained with KDD16 and several different pulse shapes at sequence order $N=8$. We include the results of the phase-randomized averaged RAP variant for comparison. Two main spurious harmonic peaks are visible in these background spectra for KDD16-RAP ($N=8$) -- at $320$ and $470$~kHz. The lower panel shows the zoomed-in view of the differential signal, $\Delta S = S_{\text{test on}}-S_{\text{test off}}$, around the $470$~kHz spurious-peak region in the presence of an artificial signal with a frequency of $470$~kHz (see text). We used the KDD16-RAP sequence at $N=8$ and applied different test field amplitudes, by varying the power amplification between $-50$ to $-10$~dBm. A sweep with range between $468$ to $472$ kHz in $0.1$ kHz steps is shown. As the test-signal amplitude increases, the dip at the $470$~kHz frequency becomes progressively more pronounced, as expected from theory, with the highest power level, $-10$~dBm, exhibiting the deepest drop. Additional peaks appear in the vicinity of the target signal frequency with higher signal amplitudes. }
    \label{fig:14}
\end{figure} 

\subsection{Frequency Sensing}
As a next step, we perform frequency-sensitive measurements to detect the frequency of the applied field. We note that CPMG has typically the optimal frequency resolution $\sim 1/T_{1,\rho}$. We perform frequency sensing with several DD methods: XY8(N), KDD16-standard, and KDD16-RAPs($N$). %The general pulse sequence is defined by: $[\text{Initilaziation}-\pi/2-(\tau/2-\pi-\tau/2)^\widetilde{N}-\pi/2-\text{Initilaziation}]$ where $\widetilde{N}$ is the number of the applied spin inversion pulses. 
The time between the centers of the pulses $\tau=T_{\text{pulse}}+\tau_{\text{ip}}$, is swept by keeping $T_{\text{pulse}}=1\,\mu$s and varying the inter-pulse separation $\tau_{\text{ip}}$ to tune the filter function's peak across the frequency range of interest, mapped to probe frequency via $f_p=1/(2\tau)$. As in all our measurements, we employed an alternating protocol, where we varied the phase of the final $\pi/2$ pulse to map the coherence onto the $|0\rangle$ or $|-1\rangle$ state. We then take the difference between the two fluorescence signals to minimize the effect of noise and the readout laser power fluctuations.

We scanned a broad frequency range to measure the applicability of the protocols for frequency sensing. To do this, we introduced an external AC signal with a known frequency, and observed resonance at the expected test frequency in each case (see main text).
To characterize the response more closely, we performed a series of frequency scans with and without any external signal. The response was quantified using the differential signal, $\Delta S = S_{\text{test on}}-S_{\text{test off}}$, where \text{test on} (\text{test off}) represents the obtained signal with (without) an applied external target field. The measured signal without an external target field is labelled a background signal. 

In the first measurement, we performed a scan without an external signal (see. the upper panel of Fig. \ref{fig:14}). In this particular measurement with KDD16 ($N=8$), the background signal exhibited spurious peaks at $320$ and $470$ kHz for the RAP pulses, which we obtained as we varied $f_p=1/(2\tau)$ by changing $\tau$ in a range from $300$ to $500$ kHz. We note that, the spurious peaks were largely absent for standard and sine-shaped pulses , which were much shorter than the RAP pulses. 
To further investigate the origin of these peaks, we applied randomization of the pulse phases in the sensing scheme~\cite{wang2019randomization}. Since a single run with randomized phases constitutes a random walk, we averaged over
multiple independent experimental runs. Despite averaging over $30$ random-phase realizations, the spurious peaks remained
present in the spectrum, indicating that this response does not originate from residual, randomizable pulse-phase errors.
Instead, this persistence is consistent with the interpretation given in the main text: the spurious harmonics likely arise from the KDD16-RAP filter function itself, as the RAP pulses are non-instantaneous, an effect that phase randomization could not suppress enough in the experiment. 
We then performed a second measurement where we added an external field with a frequency that coincided with one of these peaks, at $470$ kHz. Our goal was to investigate how the signal changes at different amplitudes of the applied field in the presence of a spurious harmonics peak at the same frequency. As shown in the lower pannel of Fig.~\ref{fig:14}, the differential signal $\Delta_\text{signal}$ exhibits a pronounced dip centered at $470$ kHz, which deepens with increasing external-field amplitude, as expected from theory. However, there are also additional peaks that appear in the vicinity of the target signal frequency with higher signal amplitudes. This shows that the differential signal measurement can in principle be used for frequency sensing even in the presence of such spurious harmonics although the obtained signal might be broader. Finally, we note the application of tailored spurious response suppression sequences, e.g., using correlated phases \cite{Wang2020correlatedPhases}, is likely to reduce this effect but this goes beyond the scope of the present work.

\end{document}